\documentclass[]{aastex631}
\usepackage{amsmath}
\usepackage{subfigure}
\usepackage{appendix}
\usepackage{CJKutf8}
\usepackage[utf8]{inputenc}
\usepackage{booktabs}
\usepackage{longtable}
\usepackage{graphicx}
\usepackage{xcolor}

\definecolor{mygray}{gray}{.9}
\shorttitle{Understanding the Evolutionary Role of star-forming S0s}
\shortauthors{Chen et al}
\graphicspath{{./}{figures/}}

\begin{document}
\begin{CJK*}{UTF8}{gbsn}


\title{Toward Understanding the Evolutionary Role of Star-forming Lenticular Galaxies: New HI Detections and Comparison with Quiescent S0s and Red Spirals}

\correspondingauthor{Junfeng Wang}
\email{jfwang@xmu.edu.cn}

\author[0009-0002-4905-263X]{Pei-bin Chen (陈培彬)}
\affiliation{Department of Astronomy, Xiamen University, Xiamen, Fujian 361005, China}

\author[0000-0003-4874-0369]{Junfeng Wang (王俊峰)}
\affiliation{Department of Astronomy, Xiamen University, Xiamen, Fujian 361005, China}

\author[0000-0002-1335-6212]{Tian-wen Cao (曹天文)}
\affiliation{Department of Astronomy, Xiamen University, Xiamen, Fujian 361005, China}

\author[0009-0003-2072-1200]{Mengting Shen (沈梦婷)}
\affiliation{Department of Astronomy, Xiamen University, Xiamen, Fujian 361005, China}

\author[0000-0003-0970-535X]{Xiaoyu Xu (许啸宇)}
\affiliation{School of Astronomy and Space Science, Nanjing University, Nanjing 210023, China}
\affiliation{Key Laboratory of Modern Astronomy and Astrophysics, Nanjing University, Nanjing 210023, China}


\begin{abstract}

As one type of blue early-type galaxies, the evolutionary history and fate of star-forming lenticular galaxies (S0s) remain elusive. We selected 134 star-forming S0s from the SDSS-IV MaNGA survey and found that they have steep and warped size-mass relations, similar to quiescent S0s and red spirals, indicating that they may have similar gas dissipation scenarios. These galaxies have a higher central stellar mass surface density than normal blue spirals. The radial profiles of $D_{\rm n}4000$ and [Mgb/Fe] show that red spirals and quiescent S0s have similar old central populations and high [Mgb/Fe] values, suggesting rapid bulge formation, though red spirals exhibit a steeper gradient possibly due to residual star formation (SF) in outer regions. In contrast, star-forming S0s exhibit profiles between quiescent S0s/red spirals and normal blue spirals, with relatively flat $D_{\rm n}4000$ and [Mgb/Fe] gradients. More long-term SF history causes normal blue spirals to have very flat $D_{\rm n}4000$ and [Mgb/Fe] profiles, and the majority of them (79 $\pm$ 5 $\%$) have S$\acute{\rm e}$rsic index $<$ 2. We also found that the halo mass of star-forming S0s resembles that of quiescent S0s/red spirals, with 82 $\pm$ 5 $\%$ exceeding the critical mass ($M_{\rm halo} = 10^{12}$$M_{\odot}$h$^{-1}$). To supplement previous H\,{\sc i} detection of star-forming S0s covered by H\,{\sc i}MaNGA, we obtained new observation for H\,{\sc i} emission from 41 star-forming S0s in our sample using the Five-hundred-meter Aperture Spherical Radio Telescope. We found that the H\,{\sc i} mass distribution of star-forming S0s matches that of normal blue spirals, although both star-forming S0s and red spirals are relatively gas-poor, resulting in varying atomic gas depletion times due to different SF levels. Based on these observational results, we discuss the possible evolutionary scenarios of star-forming S0s.

\end{abstract}


\section{Introduction}\label{intro}

Galaxies can primarily be classified based on their morphology into three main types: elliptical galaxies, spiral galaxies, and the bridge between them - lenticular galaxies \citep[S0s,][]{1926ApJ....64..321H}. These classifications correspond to two major categories in the local universe: early-type galaxies (ETGs) and late-type galaxies (LTGs). Various physical parameters of galaxies (e.g., optical colors, star formation rate, luminosity) are closely related to their morphology, and display the bimodal distribution \citep[e.g.,][]{2004ApJ...600..681B, 2013pss6.book..207V, 2014MNRAS.440..889S, 2017MNRAS.471.2687B, 2020ApJ...897..162G, 2023RAA....23a5005C}. For instance, in the color-magnitude and color-mass diagrams, ETGs primarily reside in the red sequence, while LTGs are predominantly found in the blue cloud (star-forming galaxies, SFGs), with their transition zone referred to as the green valley \citep[green-valley galaxies, GVGs;][]{2001AJ....122.1861S, 2014MNRAS.440..889S}. The evolution of galaxies from the blue cloud to the red sequence has been an active area of research \citep[e.g.,][]{2004ApJ...600..681B, 2007ApJ...665..265F, 2014MNRAS.440..889S}. Various mechanisms have been proposed to understand these issues related to this transition, such as kinematic heating \citep[e.g.,][]{2017MNRAS.465...32B}, halo quenching \citep[][]{2006MNRAS.368....2D}, active galactic nucleus (AGN) feedback \citep[e.g.,][]{2012ARA&A..50..455F}, and morphological quenching \citep[e.g.,][]{2009ApJ...707..250M, 2014MNRAS.441..599B}. \citet{2014MNRAS.441..599B} found that the bulge mass is the dominant factor of the passive galaxy fraction, indicating the importance of morphological quenching. Moreover, it is expected that galaxies hosted by dark matter halos above some critical halo mass are unable to form new stars owing to the shock heating of circumgalactic gas \citep[i.e., halo quenching;][]{2006MNRAS.368....2D}.

The majority of proposed mechanisms aim to explain gas cooling, outflow, and inflow, which have been studied intensively with new observational facilities. As the dominant component of cold gas in the universe, atomic hydrogen (H\,{\sc i}) plays a crucial role in regulating the star formation \citep[SF; e.g.,][]{2012ApJ...759....9K}. While molecular hydrogen (H$_2$) serves as the direct material for the SF, it is a non-polar molecule without an electric dipole moment, necessitating indirect measurement \citep[e.g.,][]{1991ARA&A..29..581Y, 2005ARA&A..43..677S, 2021MNRAS.504.2360J}. In contrast, the more abundant H\,{\sc i} in the local universe can be easily observed through the 21 cm hyperfine structure emission line \citep[e.g.,][]{2001MNRAS.322..486B, 2011AJ....142..170H, 2018ApJ...861...49H, 2021ApJ...918...53G}. It is widely believed that the quenching of SF is closely related to the decrease in H\,{\sc i} content \citep[e.g.,][]{2024ApJ...963...86L}. Numerous studies have identified a strong correlation between H\,{\sc i} content and the optical/UV characteristics of galaxies, revealing that galaxies with a higher H\,{\sc i} gas fraction ($f_{\rm HI}$ = $M_{\rm HI}$/$M_{\ast}$) tend to be bluer, exhibit higher star formation rates (SFR), and have lower stellar mass ($M_{\ast}$) densities \citep[e.g.,][]{2004ApJ...611L..89K, 2015MNRAS.452.2479B, 2018MNRAS.476..875C, 2020ApJ...897..162G, 2022MNRAS.516.2337W}. However, the higher H\,{\sc i} detection rate in red spirals challenges this consensus \citep[e.g.,][]{2018ApJ...861...49H, 2020ApJ...897..162G}.

Since their discovery, red spirals have garnered significant attention \citep[e.g.,][]{2010MNRAS.405..783M, 2019ApJ...883L..36H, 2019ApJ...880..149P, 2020ApJ...897..162G}. They are outliers in the well-known morphology-density relation, raising questions about their nature, including whether they are precursors to S0s \citep[e.g.,][]{2003MNRAS.346..601G, 2009MNRAS.393.1302W, 2010MNRAS.405..783M}. The proportion of red spirals in the low-mass end is very small \citep[e.g.,][]{2010MNRAS.405..783M, 2018MNRAS.474.1909F, 2019ApJ...880..149P}, leading most studies to focus on massive red spirals \citep[log $M_{\ast}$ $>$ 10.5 $M_{\odot}$; e.g.,][]{2019ApJ...883L..36H, 2020ApJ...897..162G}. \citet{2010MNRAS.405..783M} used Galaxy Zoo to construct a sample of optically ($g - r$) defined red spirals, representing a compelling set of potential transition objects between normal blue spirals and red ETGs, accounting for 6$\%$ of LTGs. They proposed several possible origins for red spirals: they may be old spirals that have used up fuel, the transformation of normal blue spirals that have undergone some processes \citep[ram pressure stripping, or starvation/strangulation; e.g.,][]{1972ApJ...176....1G, 2022A&A...659A..46B, 2022MNRAS.515.5877K}, or the evolution of normal blue spirals due to bar instability. \citet{2019ApJ...883L..36H} and \citet{2020ApJ...897..162G} studied the optically ($u - r$) defined massive red spirals, emphasizing that interactions or mergers are crucial to the formation and evolution of these galaxies and that morphology, halo, and angular momentum quenching collectively cease their SF. Specifically, \citet{2019ApJ...880..149P} used $NUV - r$ to select 9 passive spirals from the CALIFA survey, suggesting that they may be one of the channels for the formation of S0s. More recently, \citet{2024MNRAS.528.2391C} used the SDSS-IV MaNGA survey to study the structures and stellar populations (SPs) characteristics of $g - r$ massive red spirals. Their findings indicate that the similarities between massive red spirals and S0s reflect a potential evolutionary trend applicable to all S0s across different environments. They suggested that massive red spirals may evolve into S0s as their spiral arms fade and residual SF is exhausted \citep[][]{2024MNRAS.528.2391C}.

Previous studies have suggested a possible evolutionary trend between these galaxies: normal blue spirals undergo various processes to transform into red spirals \citep[][]{2010MNRAS.405..783M}, which evolve into quiescent S0s as their less pronounced spiral arms fade and residual SF is exhausted \citep[][]{2024MNRAS.528.2391C}. While it is confirmed that S0s have complex and diverse formation pathways, they can generally be categorized into two main types \citep[e.g.,][]{2020MNRAS.492.2955C, 2020MNRAS.498.2372D, 2021MNRAS.508..895D, 2022MNRAS.509.1237X, 2024A&A...691A.107C}. Based on the similar disk structure and kinematics, S0s are commonly believed to have originated from spirals, often referred to as $faded\ spirals$ \citep[][]{1972ApJ...176....1G, 2009MNRAS.394.1991B, 2012ApJS..198....2K, 2018MNRAS.478..351M, 2020MNRAS.492.2955C, 2020MNRAS.498.2372D, 2021MNRAS.508..895D, 2024A&A...691A.107C}. In contrast, S0s found in low-density environments display significantly different characteristics (e.g., redder bulges), and the observed kinematic decoupling of gas suggests external origins for this gas. These galaxies are generally thought to result from $minor\ mergers$ \citep[][]{2009ApJ...700.1702V, 2018MNRAS.477.2030D, 2020MNRAS.492.2955C, 2020MNRAS.498.2372D, 2021MNRAS.508..895D, 2024A&A...691A.107C}. Although SF has ceased in most S0s, recent studies indicate that some S0s continue to exhibit ongoing SF \citep[e.g.,][]{2015A&A...583A.103G, 2022MNRAS.513..389R, 2022MNRAS.509.1237X, 2023RAA....23a5005C, 2023A&A...678A..10G, 2024A&A...691A.107C}, suggesting that star-forming S0s may have undergone different physical processes during their formation and evolution compared to quiescent S0s. 

Star-forming S0s, as part of the category of star-forming blue ETGs (star-forming ellipticals/S0s), have garnered significant attention \citep[e.g.,][]{2015A&A...583A.103G, 2022MNRAS.513..389R, 2022MNRAS.509.1237X, 2023RAA....23a5005C, 2023A&A...678A..10G, 2024A&A...691A.107C}. However, the position of star-forming blue ETGs in the possible evolutionary sequence remains confusing \citep[][]{2009AJ....138..579K, 2009MNRAS.396..818S}. \citet{2009MNRAS.396..818S} utilized the Galaxy Zoo to identify 204 blue ETGs, discussing their possible role in the overall evolutionary framework. They suggested that these galaxies might result from spiral-spiral mergers, evolving into red-sequence galaxies through passive evolution. Alternatively, they could be ETGs experiencing an episode of SF due to the sudden availability of cold gas, or result from early-type/late-type mergers \citep[e.g.,][]{2003ApJ...597L.117K}. \citet{2015A&A...583A.103G} undertook an optical and ultraviolet (UV) study of 55 star-forming blue ETGs, searching for signatures of recent interactions that may drive gas into the galaxy and trigger the SF. They suggested that recent or ongoing interactions with gas-rich neighboring galaxies may prompt cold gas to enter passively evolving ETGs, and the sudden supply of the gas triggers SF, turning them into blue ETGs. This was further confirmed by \citet{2023A&A...678A..10G}. Such an evolutionary sequence is a manifestation of how SF in galaxies quenches and rejuvenates. It is generally believed that the quenching of galaxies is related to their H\,{\sc i} content \citep[e.g.,][]{2012ApJ...759....9K, 2021ApJ...918...53G, 2023A&A...678A..10G}. Understanding H\,{\sc i} in specific galaxies can shed light on the physical processes underlying gas-driven SF and quenching \citep[e.g.,][]{2023MNRAS.523.1140L, 2024ApJ...963...86L}. Therefore, investigating the H\,{\sc i} emission of star-forming S0s — one of the components of star-forming blue ETGs — and their role in evolutionary scenarios is particularly intriguing.

In this study, we selected 134 star-forming S0s from the SDSS-IV MaNGA survey and aimed to explore their evolutionary scenarios using H\,{\sc i} emission in conjunction with other data, such as global structures \citep[e.g.,][]{2021ApJ...918...53G, 2024Natur.632.1009W}. We found that only 15 of these sources were included in the H\,{\sc i}MaNGA catalog \citep[][]{2019MNRAS.488.3396M, 2021MNRAS.503.1345S}. As a supplement to the H\,{\sc i} detection of star-forming S0s covered by H\,{\sc i}MaNGA, we obtained new observations of H\,{\sc i} emission from 41 star-forming S0s in our sample using the Five-hundred-meter Aperture Spherical Radio Telescope (FAST). Among these, 11 star-forming S0s exhibited a signal-to-noise ratio (S/N) greater than 5, yielding a detection rate of 27$\%$. We then compared the global structures, SPs properties, galaxy halo masses, and H,{\sc i} content of star-forming S0s with those of normal blue spirals, optically ($g - r$) defined red spirals, and quiescent S0s to discuss their possible evolutionary scenarios.

The organizational structure of this paper is as follows. In Section \ref{data}, we briefly introduce the selection of star-forming S0s and control samples, followed by a detailed description of the FAST spectra processing to obtain parameters for the H\,{\sc i} spectra. The results and discussion are presented in Sections \ref{result} and \ref{discussion}. In Section \ref{conclusions}, we provide relevant conclusions. Throughout this paper, we adopt a set of cosmological parameters as follows: \emph{H$_0$} = 70 km s$^{-1}$ Mpc$^{-1}$ (i.e., \emph{h} = 0.7), $\Omega_{\rm m}$ = 0.30, and $\Omega_{\rm \Lambda}$ = 0.70.

\section{Sample and Observation} 
\label{data}
\subsection{Sample Selection} \label{sec2.1}

MaNGA is a component of the SDSS-IV \citep[][]{2015ApJ...798....7B, 2017AJ....154...28B}, whose goal is to map the detailed composition and kinematic structures of nearby galaxies. It uses integral field unit (IFU) spectroscopy to measure spectra for hundreds of points within each galaxy. The IFUs are used to spectroscopically map galaxies over the wavelength range of 3600\,\AA\ to 10400\,\AA\ at a resolution of roughly 2000 \citep[$\lambda$/$\delta\lambda$;][]{2015ApJ...798....7B}. In this work, we selected the targets from the deep-learning catalog of galaxy morphology \citep[hereafter, MDLM-VAC\footnote{\url{https://www.sdss.org/dr17/data_access/value-added-catalogs/}\label{deep_learning_cata}};][]{2022MNRAS.509.4024D}, and the steps are as follows:

1). We selected galaxies that pass the basic selection criteria and have $T\_Type \leq 0$, $P\_LTG < 0.5$, $P\_S0 > 0.5$, $VC = 2$ as recommended by MDLM-VAC for identifying S0s. We obtained 924 S0s candidates (hereafter, S0s$\_$candidates). Following the method provided by \citet{2022MNRAS.512.2222V}, MaNGA provides a pure visual morphology classification catalog (hereafter, MaNGA-visual-morpho), covering all galaxies in NaNGA Data Release 17 (DR17). This classification is derived from the inspection of image mosaics utilizing a new re-processing of SDSS and Dark Energy Legacy Survey (DESI) images. Digital image processing utilizes the advantages of both SDSS and DESI images, facilitating the identification of internal structures and low surface brightness features. To ensure the correct classification of S0, we cross-matched S0s$\_$candidates with MaNGA-visual-morpho and then selected galaxies that satisfy $Hubble-type = S0$ (906 targets). Finally, the unreliable classifications were excluded (i.e., $Unsure = 1$), obtaining 814 galaxies classified as S0.

2). We performed a global Baldwin, Phillips $\&$ Telervich \citep[BPT,][]{1981PASP...93....5B} diagnosis of the galaxy using the emission line flux fitted within 1 effective radius ($R_{\rm e}$) from Pipe3D catalog\footnote{\url{https://data.sdss.org/datamodel/files/MANGA_PIPE3D/MANGADRP_VER/PIPE3D_VER/SDSS17Pipe3D.html}} \citep[][]{2020ARA&A..58...99S}. This diagnosis helped us to select galaxies located in the star-forming region of the BPT diagram, and the theoretical boundaries of the BPT diagram come from \cite{2003MNRAS.346.1055K}, \cite{2001ApJ...556..121K}, and \cite{2007MNRAS.382.1415S}, respectively. Then, we further required that the H$_{\alpha}$ equivalent width (EW$\_$H$_{\alpha}$) of the galaxy $>$ 6\,\AA\ within 2.5$^{\prime \prime}$ at the center to ensure it is SFGs \citep[e.g.,][]{2020ARA&A..58...99S, 2022MNRAS.509.1237X}. Finally, we obtained a sample containing 134 star-forming S0s. Note that in the selection process of S0s, we did not restrict the axis ratio (b/a) of the galaxy, because if we truncated according to the b/a used in previous work \citep[$>$0.3; e.g.,][]{2022MNRAS.509.1237X, 2024A&A...691A.107C}, we found that the final star-forming S0s only reduced by about 4$\%$, which would not have a significant impact on our results. The most important thing is that there is almost no detection of H\,{\sc i} in the reduced galaxies.

\subsection{Comparison Samples}\label{sec2.2}

For comparison with the above 134 star-forming S0s, we also selected optically ($g - r$) defined red spirals, quiescent S0s, and normal blue spirals from MDLM-VAC. \citet{2024MNRAS.528.2391C} selected a group of massive red spirals from the SDSS-IV MaNGA survey using the dust-corrected $g - r$ criterion combined with WISE color. In this paper, we follow similar selection criteria as \citet{2024MNRAS.528.2391C} to select red spirals from the SDSS-IV MaNGA survey. Our steps are as follows: 

1). We selected galaxies meeting the criteria $T\_Type > 0$, $P\_LTG \geq 0.5$ and $VC = 3$ from MDLM-VAC, and obtained a parent sample that only contains spiral galaxies (hereafter, SGs$\_$parent sample, 3997 galaxies). Then, we cross-matched SGs$\_$parent sample with NYU Value-Added Galaxy catalog\footnote{\url{http://sdss.physics.nyu.edu/vagc/}} \citep[NYU$\_$VAGC;][]{2005AJ....129.2562B} to obtain the magnitude of the g-band and r-band after dust correction (3602 galaxies).

\begin{figure}[!ht]
 \centering
 \subfigure[]
 {
  \begin{minipage}{8cm}
   \centering
   \includegraphics[scale=0.45]{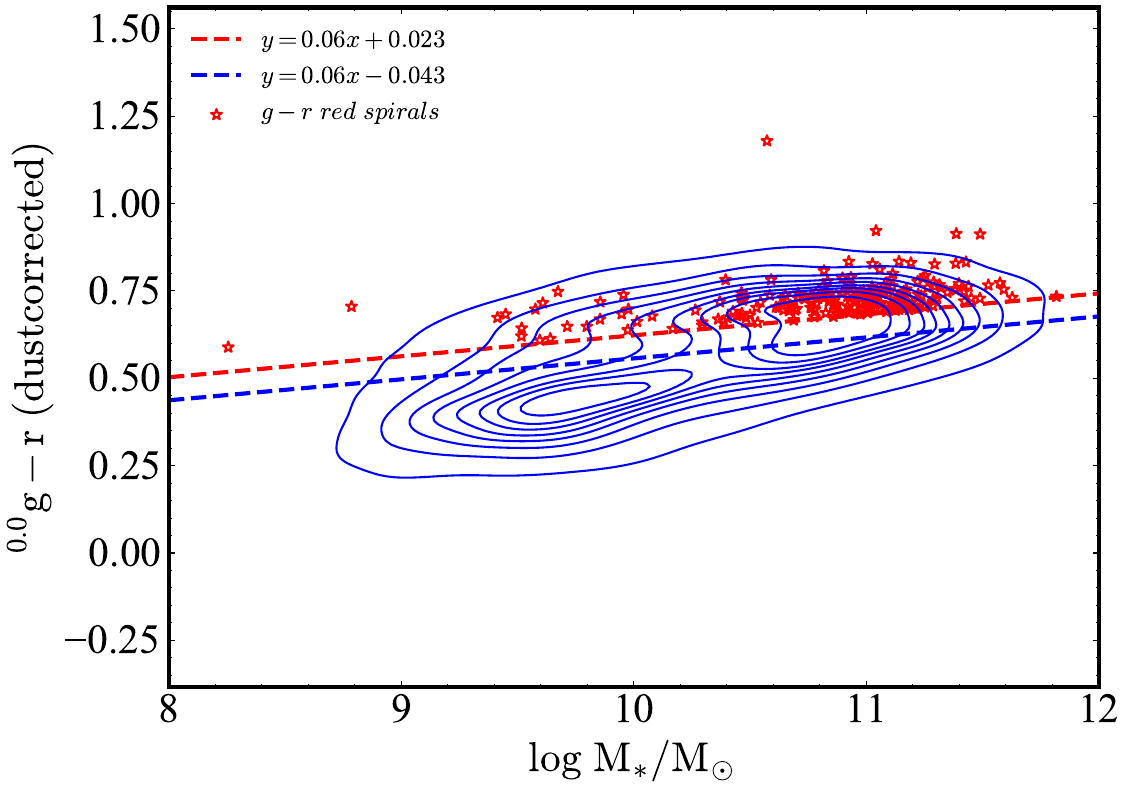}
   \label{gr_color}
  \end{minipage}
 }
    \subfigure[]
    {
     \begin{minipage}{8cm}
      \centering
      \includegraphics[scale=0.45]{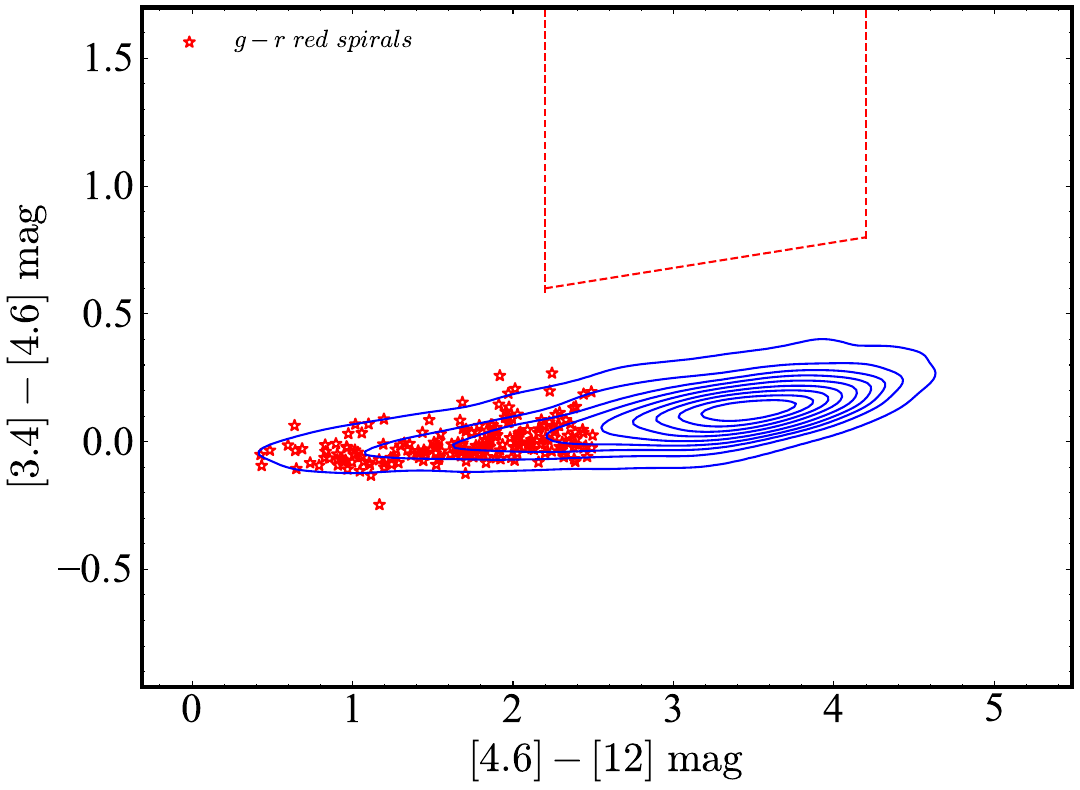}
      \label{wise_color}
     \end{minipage}
    }
\caption{Distributions of red spirals and SGs$\_$parent sample in color-mass and color-color diagrams. a) The $g - r$ color-mass diagram of red spirals. The dashed lines show the separation between the red-sequence galaxies, GVGs, and SFGs, defined as $g - r = 0.06 \times {{\rm log}M_{\ast}/M_{\odot}} + 0.023$ (red) and $g - r = 0.06 \times {{\rm log}M_{\ast}/M_{\odot}} - 0.043$ (blue) respectively \citep[][]{2024MNRAS.528.2391C}. b) The WISE color-color diagram of $g - r$ red spirals. The AGN region is defined by \citet{2011ApJ...735..112J}, shown as the red dashed lines. In this Figure, the open red stars represent $g - r$ red spirals and the blue contours show the number density distribution of our SGs$\_$parent sample.}
\label{color_dis}
\end{figure}

2). Then, we selected the red spirals based on the $g - r$ color criterion \citep[$g - r > 0.06 \times {{\rm log}M_{\ast}/M_{\odot}} + 0.023$;][]{2024MNRAS.528.2391C} and the WISE\footnote{\url{https://irsa.ipac.caltech.edu/cgi-bin/Gator/nph-dd}} \citep[][]{2010AJ....140.1868W} color \citep[WISE2 - WISE3 $<$ 2.5;][]{2024MNRAS.528.2391C}. To reduce the influence of internal dust extinction, we required the axis ratio (b/a) of the galaxy $\geq$ 0.5. Finally, we obtained a sample containing 214 $g - r$ red spirals. Fig. \ref{color_dis} shows the color-mass and WISE color-color diagrams, where the open red stars are the $g - r$ red spirals. In this figure, the blue contours show the number density distribution of our SGs$\_$parent sample. In Fig. \ref{gr_color}, dashed lines represent the separation between red-sequence galaxies, GVGs, and SFGs, defined as $g - r = 0.06 \times {{\rm log}M_{\ast}/M_{\odot}} + 0.023$ (red) and $g - r = 0.06 \times {{\rm log}M_{\ast}/M_{\odot}} - 0.043$ (blue) respectively \citep[][]{2024MNRAS.528.2391C}. The red dashed lines in Fig. \ref{wise_color} represent the region occupied by the AGN host galaxies \citep[][]{2011ApJ...735..112J}.

Moreover, to compare with them, we selected normal blue spirals from the SGs$\_$parent sample. Briefly, we conducted a global BPT diagnosis, selecting galaxies located in the H\,{\sc ii} regions and requiring their EW$\_$H$_{\alpha}$ $>$ 6\,\AA\ within the center 2.5$^{\prime \prime}$. Finally, we obtained 2642 normal blue spirals. Similar to the steps in Sect. \ref{sec2.1}, we also obtained 680 quiescent S0s. The other galaxy parameters (e.g., $R_{\rm e}$, b/a, S$\acute{\rm e}$rsic index) used in our work are from the NASA-Sloan Atlas (NSA) catalog\footnote{\url{https://data.sdss.org/datamodel/files/ATLAS DATA/ATLAS MAJOR VERSION/nsa.html}} \citep[v1$\_$0$\_$1,][]{2011AJ....142...31B}, while SFR and $M_{\ast}$ are sourced from the fitting catalog of spectral energy distribution (SED) by \cite{2016ApJS..227....2S, 2018ApJ...859...11S}. The archival H\,{\sc i} data come from H\,{\sc i}MaNGA \citep[][]{2019MNRAS.488.3396M, 2021MNRAS.503.1345S}. Our four samples are listed in Table \ref{basic information}.

\begin{deluxetable}{ccccccccc}
\centering
\tablenum{1}
\small
\tablecaption{The information of our samples\label{basic information}}
\tablehead{
\begin{tabular}[c]{@{}c@{}}Samples\\ (1)\end{tabular} & \begin{tabular}[c]{@{}c@{}}Size\\ (2)\end{tabular} & \begin{tabular}[c]{@{}c@{}}H\,{\sc i} Detection\\ (3)\end{tabular} & \colhead{\begin{tabular}[c]{@{}c@{}}S$\acute{\rm e}$rsic index\\(median)\\ (4)\end{tabular}} & \colhead{\begin{tabular}[c]{@{}c@{}}S$\acute{\rm e}$rsic index $<$ 2\\ (5)\end{tabular}} & \begin{tabular}[c]{@{}c@{}}log $\Sigma_{1}$\\ ($M_{\odot}$ kpc$^{-2}$)\\(median)\\ (6)\end{tabular} & \begin{tabular}[c]{@{}c@{}}log $\Sigma_{1}$ $>$ 9.4\\ ($M_{\odot}$ kpc$^{-2}$)\\ (7)\end{tabular} & \begin{tabular}[c]{@{}c@{}}log $M_{\rm halo}$\\ ($M_{\odot}$h$^{-1}$)\\(median)\\ (8)\end{tabular} & \begin{tabular}[c]{@{}c@{}}log $M_{\rm halo}$ $>$ 12\\ ($M_{\odot}$h$^{-1}$)\\ (9)\end{tabular}
}
\startdata
Normal blue spirals & 2642 (2332) & 1381 (923) & 1.31$\pm$0.32 & 79$\pm$5$\%$ & 8.46$\pm$0.13 & $\sim$2.35$\%$ & 12.20$\pm$0.10 & 65$\pm$4$\%$ \\
$g - r$ red spirals & 214 (159) & 64 (43) & 4.42$\pm$0.70 & 9$\pm$2$\%$ & 9.37$\pm$0.06 & 47$\pm$5$\%$ & 12.74$\pm$0.12 & 92$\pm$7$\%$ \\
Quiescent S0s & 680 & - - & 4.15$\pm$0.64 & 4$\pm$2$\%$ & 9.19$\pm$0.07 & 22$\pm$3$\%$ & 12.66$\pm$0.16 & 83$\pm$5$\%$ \\
star-forming S0s & 134 (112) & 15 (3) & 3.61$\pm$0.64 & 13$\pm$2$\%$ & 8.91$\pm$0.09 & 9$\pm$2$\%$ & 12.73$\pm$0.16 & 82$\pm$5$\%$ \\
\enddata
\tablecomments{The columns show (1) our four samples; (2) the size of samples. Format a(b) represents the total number of galaxies (a) and the number of galaxies with redshift $<$ 0.05 (b); (3) H\,{\sc i} detection (S/N $>$ 3) from H\,{\sc i}MaNGA; H\,{\sc i}MaNGA \citep[][]{2019MNRAS.488.3396M, 2021MNRAS.503.1345S} suggests that S/N $>$ 3 indicates the detection of H\,{\sc i} in the galaxy. Format a(b) represents the number of galaxies with S/N $>$ 3 (a) and the number of galaxies with S/N $>$ 5 (b); (4) Median S$\acute{\rm e}$rsic index from NSA; (5) Proportion of S$\acute{\rm e}$rsic index $<$ 2; (6) Median log $\Sigma_{1}$; (7) Proportion of log $\Sigma_{1}$ $>$ 9.4; (8) Median galaxy halo mass from \cite{2007ApJ...671..153Y}; (9) Proportion exceeding the critical mass.}
\end{deluxetable}

\subsection{HI Observations}\label{2.3}

\subsubsection{FAST Observation}\label{2.3.3}

Based on the information above and considering that S0s may lack the atomic and molecular gas, we further refined our criteria for the 134 star-forming S0s by requiring their SFR greater than 1 M$_{\odot}$ yr$^{-1}$ to ensure detection by the Five-hundred-meter Aperture Spherical Radio Telescope (FAST). This process yielded a final observational sample of 41 star-forming S0s. We conducted observations of these galaxies using FAST \citep[][]{2006ScChG..49..129N, 2019SCPMA..6259502J, 2020RAA....20...64J, 2020Innov...100053Q}. FAST is a single-dish radio telescope with an effective diameter of 300 meters, renowned for its highest sensitivity among ground-based radio telescopes. It features 19 beam receivers covering a frequency range of 1.05–1.45 GHz, with a frequency resolution of approximately 7.63 kHz and an angular resolution of about 2.9$^{\prime}$ at 1.4 GHz \citep[][]{2022MNRAS.516.2337W}. The temperature of the observation system, including the sky background, is approximately 20 Kelvin (K) with a zenith angle of 26.4 degrees. The basic information of these 41 targets is listed in Table \ref{tab1}.

Using the position switch ON-OFF mode, we observed our targets through a Shared-Risk project of FAST (project ID: PT2023$\_$0112, PI: Peibin Chen). Both polarizations (xx and yy) of all targets are recorded. Observations took place on September 3 and 4, 2023. The integration time for ON-source and OFF-source was set to 109 seconds, with a switch time of 30 seconds (overhead) between ON and OFF positions, resulting in a total integration time of 248 seconds for each source. We operated in low noise mode to reduce baseline ripples, achieving a median noise diode temperature of about 1.1 K \citep[][]{2020RAA....20...64J}. For our targets, the sampling time is 1.0 seconds, and each target was observed using the same configuration. We utilized the data reduction pipeline H\,{\sc i}FAST\footnote{\url{https://hifast.readthedocs.io/zh/v1.2/hifast.sw.html}} developed by \cite{Jing2023} based on Python to process observation data, which includes the following steps:

1). Noise Diode Separation and Calibration: Separate the noise diode (ON or OFF), as well as the source and reference point, and then perform noise diode calibration. Here, the unit of the spectrum is calibrated to K, and the calibration signal comes from a standard 1 K noise diode injected every 8 s. Subtract the reference point from the source and merge it into one spectrum, retaining two polarizations. The detailed processes are described in \cite{2022RAA....22b5015Z}.
    
2). Baseline and Standing Wave Removal: Using the asymmetrically re-weighted penalized least squares algorithm \citep[][]{2015EPJB...88..127B} to subtract the baseline from each spectrum. Then, use the $sin$ function to fit the standing wave. Notably, during the baseline removal process, it is necessary to perform on both polarizations and save them.
    
3). Flux Calibration: Calibrate the flux unit from K to Jansky (Jy) using the conversion factor provided by \cite{2019SCPMA..6259502J}.

\begin{deluxetable}{ccccccccccc}
\centering
\tablenum{2}
\tiny
\tablecaption{Information of our star-forming S0s\label{tab1}}
\tablehead{
\begin{tabular}[c]{@{}c@{}}MaNGA$\_$ID\\ (Plate$\_$IFU)\\ (1)\end{tabular} & \begin{tabular}[c]{@{}c@{}}R.A.\\ (deg.)\\ (2)\end{tabular} & \begin{tabular}[c]{@{}c@{}}Decl.\\ (deg.)\\ (3)\end{tabular} & \colhead{\begin{tabular}[c]{@{}c@{}}Redshift\\ ($z$)\\ (4)\end{tabular}} & \begin{tabular}[c]{@{}c@{}}log M$_{\ast}$\\ ($M_{\odot}$)\\ (5)\end{tabular} & \begin{tabular}[c]{@{}c@{}}log SFR$_{\rm SED}$\\ ($M_{\odot}$yr$^{-1}$)\\ (6)\end{tabular} & \begin{tabular}[c]{@{}c@{}}W$_{50}$\\ (km s$^{-1}$)\\ (7)\end{tabular} & \begin{tabular}[c]{@{}c@{}}log M$_{\rm HI}$\\ ($M_{\odot}$)\\ (8)\end{tabular} & \begin{tabular}[c]{@{}c@{}}$S_{21}$\\ (Jy km s$^{-1}$)\\ (9)\end{tabular} & \begin{tabular}[c]{@{}c@{}}S/N\\ (10)\end{tabular} & \begin{tabular}[c]{@{}c@{}}rms\\ (mJy)\\ (11)\end{tabular}
}
\startdata
9183-12705 & 123.10611 & 37.73022 & 0.038 & 10.487 $\pm$ 0.036 & 0.592 $\pm$ 0.08 &  & 9.34 & 0.36 & & 1.6\\
8252-3704 & 145.30812 & 47.68860 & 0.047 & 10.526 $\pm$ 0.041 & 0.520 $\pm$ 0.147 &  & 9.67 & 0.41 & & 1.9\\
8568-9101 & 155.58076 & 36.58304 & 0.026 & 10.948 $\pm$ 0.031 & 0.527 $\pm$ 0.114 &  & 9.28 & 0.68 &   & 3.0\\
8441-1901 & 222.86356 & 37.50326 & 0.032 & 10.227 $\pm$ 0.050 & 0.185 $\pm$ 0.207 &  & 9.31 & 0.48 & & 2.0\\
8723-6104 & 130.40784 & 54.91850 & 0.045 & 10.640 $\pm$ 0.034 & 0.493 $\pm$ 0.087 &  & 9.68 & 0.58 & & 2.0\\
\hline
8720-3701 & 120.83615 & 48.58843 & 0.058 & 10.960 $\pm$ 0.045 & 0.178 $\pm$ 0.382 &  & 9.74 & 0.40 &  & 1.8 \\
8715-6101 & 119.10579 & 51.34457 & 0.054 & 10.789 $\pm$ 0.037 & 0.131 $\pm$ 0.181 &  & 9.72 & 0.44 & & 2.0\\
7993-3701 & 33.62695 & 13.25721 & 0.060 & 11.213 $\pm$ 0.048 & 0.591 $\pm$ 0.152 &  & 9.78 & 0.41 & & 1.8\\
9091-3704 & 241.94466 & 25.53750 & 0.041 & 10.081 $\pm$ 0.059 & 0.224 $\pm$ 0.099 & 182.70 $\pm$ 4.74 & 9.53 $\pm$ 0.04 & 0.50 $\pm$ 0.04 & 13.00 & 0.6\\
9498-3702 & 118.34623 & 24.55016 & 0.061 & 10.412 $\pm$ 0.034 & 0.705 $\pm$ 0.046 & 225.30 $\pm$ 10.84 & 9.63 $\pm$ 0.05 & 0.28 $\pm$ 0.04 & 6.25 & 0.7\\
12094-3701 & 18.19596 & 14.36086 & 0.057 & 10.787 $\pm$ 0.031 & 0.437 $\pm$ 0.146 &  & 9.91 & 0.62 &  & 2.8\\
11960-3701 & 236.26817 & 8.55968 & 0.042 & 10.066 $\pm$ 0.042 & 0.045 $\pm$ 0.121 &  & 9.43 & 0.37 & & 1.7\\
11976-3701 & 241.56182 & 18.18289 & 0.039 & 10.884 $\pm$ 0.056 & 0.808 $\pm$ 0.119 & 111.90 $\pm$ 4.32 & 9.48 $\pm$ 0.05 & 0.48 $\pm$ 0.04 & 13.13 & 0.8\\
12090-6101 & 350.86673 & 14.09174 & 0.041 & 10.314 $\pm$ 0.041 & 0.299 $\pm$ 0.052 & 97.87 $\pm$ 10.72 & 9.14 $\pm$ 0.05 & 0.19 $\pm$ 0.03 & 6.74 & 0.7\\
12685-3704 & 329.54187 & $-7.80036$ & 0.057 & 10.836 $\pm$ 0.047 & 0.086 $\pm$ 0.341 &  & 9.94 & 0.66 & & 3.0\\
9196-3702 & 261.92788 & 54.05235 & 0.080 & 11.194 $\pm$ 0.034 & 0.509 $\pm$ 0.174 &  & 9.60 & 0.16 & & 0.7\\
8262-3703 & 184.84971 & 44.04161 & 0.067 & 11.130 $\pm$ 0.020 & 0.410 $\pm$ 0.096 &  & 9.94 & 0.48 &  & 2.0\\
8323-3703 & 196.43983 & 34.68108 & 0.067 & 11.065 $\pm$ 0.027 & 0.557 $\pm$ 0.099 &  & 9.31 & 0.11 & & 2.5\\
8250-3703 & 139.73996 & 43.50058 & 0.040 & 9.723 $\pm$ 0.045 & 0.434 $\pm$ 0.071 & 86.16 $\pm$ 4.73 & 9.28 $\pm$ 0.05 & 0.29 $\pm$ 0.03 & 10.52 & 0.7\\
8996-3704 & 173.41287 & 52.67459 & 0.049 & 10.266 $\pm$ 0.031 & 0.027 $\pm$ 0.214 &  & 9.80 & 0.65 & & 2.9\\
8618-3704 & 318.86229 & 9.75782 & 0.070 & 10.978 $\pm$ 0.053 & 0.788 $\pm$ 0.136 & & 9.89 & 0.40 & & 1.8\\
8977-6104 & 118.80956 & 33.35323 & 0.082 & 11.195 $\pm$ 0.039 & 1.007 $\pm$ 0.091 &  & 10.04 & 0.40 & & 1.8\\
9500-6104 & 132.17901 & 26.02348 & 0.022 & 9.672 $\pm$ 0.026 & 0.043 $\pm$ 0.023 & 218.60 $\pm$ 1.02 & 9.37 $\pm$ 0.09 & 1.176 $\pm$ 0.04 & 31.30 & 0.5\\
8241-9102 & 127.17080 & 17.58140 & 0.066 & 11.247 $\pm$ 0.049 & 0.471 $\pm$ 0.317 &  & 9.87 & 0.42 & & 1.9\\
10837-9102 & 159.34845 & 2.31265 & 0.040 & 10.750 $\pm$ 0.027 & 0.208 $\pm$ 0.182 &  & 9.97 & 1.40 & & 6.0\\
11760-1902 & 192.76854 & 48.40855 & 0.049 & 10.310 $\pm$ 0.059 & 0.516 $\pm$ 0.084 & 81.81 $\pm$ 4.87 & 9.49 $\pm$ 0.05 & 0.31 $\pm$ 0.03 & 10.47 & 0.7\\
11826-6104 & 190.00075 & 36.87242 & 0.065 & 10.922 $\pm$ 0.026 & 0.407 $\pm$ 0.103 &  & 9.87 & 0.43 & & 2.0\\
7993-1902 & 32.88983 & 13.91713 & 0.027 & 10.514 $\pm$ 0.042 & 0.822 $\pm$ 0.102 & 90.36 $\pm$ 2.10 & 9.11 $\pm$ 0.04 & 0.44 $\pm$ 0.02 & 27.32 & 0.4\\
8245-3701 & 134.94685 & 20.59527 & 0.025 & 9.802 $\pm$ 0.051 & 0.091 $\pm$ 0.109 & 132.90 $\pm$ 4.59 & 8.92 $\pm$ 0.04 & 0.31 $\pm$ 0.03 & 9.63 & 0.8\\
8546-3704 & 238.88515 & 50.47860 & 0.044 & 10.292 $\pm$ 0.041 & 0.005 $\pm$ 0.204 &  & 9.66 & 0.56 & & 2.5\\
8565-3703 & 242.71587 & 48.91090 & 0.045 & 10.172 $\pm$ 0.063 & 0.675 $\pm$ 0.076 &  & 9.52 & 0.39 & & 1.8\\
9499-9101 & 119.06976 & 26.88610 & 0.027 & 10.766 $\pm$ 0.040 & 0.189 $\pm$ 0.249 &  & 9.19 & 0.50 & & 2.0\\
9885-1901 & 239.35970 & 23.27251 & 0.023 & 9.811 $\pm$ 0.068 & 0.154 $\pm$ 0.135 & 83.77 $\pm$ 4.41 & 8.74 $\pm$ 0.05 & 0.24 $\pm$ 0.03 & 7.79 & 0.7\\
10843-1901 & 149.45641 & $-0.21094$ & 0.033 & 10.017 $\pm$ 0.039 & 0.283 $\pm$ 0.033 &  & 9.65 & 0.98 & & 4.4\\
10846-6104 & 154.52782 & 0.09992 & 0.048 & 10.947 $\pm$ 0.022 & 0.375 $\pm$ 0.092 &  & 10.13 & 1.64 & & 7.0\\
11753-3701 & 146.66127 & 2.65893 & 0.083 & 11.308 $\pm$ 0.031 & 0.943 $\pm$ 0.064 &  & 10.42 & 0.77 & & 3.0\\
11759-3701 & 145.59401 & 0.30942 & 0.046 & 10.505 $\pm$ 0.053 & 0.156 $\pm$ 0.202 &  & 9.78 & 0.70 & & 3.0\\
12087-6104 & 349.82666 & 13.91164 & 0.052 & 10.987 $\pm$ 0.040 & 0.330 $\pm$ 0.120 & & 9.82 & 0.60 &  & 3.0\\
7965-1902 & 318.50226 & 0.53510 & 0.027 & 10.237 $\pm$ 0.026 & 0.377 $\pm$ 0.036 & 59.84 $\pm$ 4.48 & 8.97 $\pm$ 0.05 & 0.31 $\pm$ 0.06 & 5.14 & 0.7\\
11974-3704 & 239.72545 & 9.52213 & 0.040 & 9.966 $\pm$ 0.021 & 0.260 $\pm$ 0.037 &  & 9.42 & 0.40 & & 1.8\\
12090-3701 & 352.61398 & 13.54854 & 0.064 & 11.152 $\pm$ 0.047 & 0.465 $\pm$ 0.190 &  & 9.83 & 0.41 & & 1.8\\
\enddata
\tablecomments{The columns show (1) the MaNGA ID of our targets; (2) R.A. in degrees; (3) Decl. in degrees; (4) redshift from NSA; (5), (6) are M$_{\ast}$ and SFR, both from \cite{2016ApJS..227....2S, 2018ApJ...859...11S}; (7), (8), (9), (10) and (11) are estimated H\,{\sc i} spectral parameters \citep[][]{2022MNRAS.516.2337W}. The numbers without error values in columns 8, 9, and 11 represent the upper limit of the estimate.}
\end{deluxetable}
    
4). Coordinate System Correction: Correct the rotation of the Earth and convert the redshift velocity from the local standard of rest to the heliocentric velocity in the equatorial coordinate system \citep[e.g.,][]{2022RAA....22f5019K}.

After reducing the original observation data using the data pipeline, we calculated key parameters (e.g., W$_{50}$ and M$_{\rm HI}$) in the H\,{\sc i} spectrum following the methods outlined in \cite{2018ApJ...861...49H} and \cite{2022MNRAS.516.2337W}. Detailed information on the calculation process can be found in Sect. 2.2 of \cite{2022MNRAS.516.2337W}. For those sources with the S/N $\leq$ 5.0, we used the same method as \cite{2022MNRAS.516.2337W} to provide the upper limit of their parameters. After processing with H\,{\sc i}FAST, we found that some galaxies in our sample have H\,{\sc i} observation frequencies near 1380 MHz. According to the statistics on radio frequency interference \citep[RFI;][]{2022RAA....22b5015Z} from FAST, this frequency is associated with a high likelihood of interference from navigation satellites\footnote{\url{https://fast.bao.ac.cn/cms/category/rfi_monitoring/}} (e.g., 8568-1901, 9499-9101). 

\begin{figure}[htbp]
 \centering
 \subfigure[]
 {
  \begin{minipage}{8cm}
   \centering
   \includegraphics[scale=0.45]{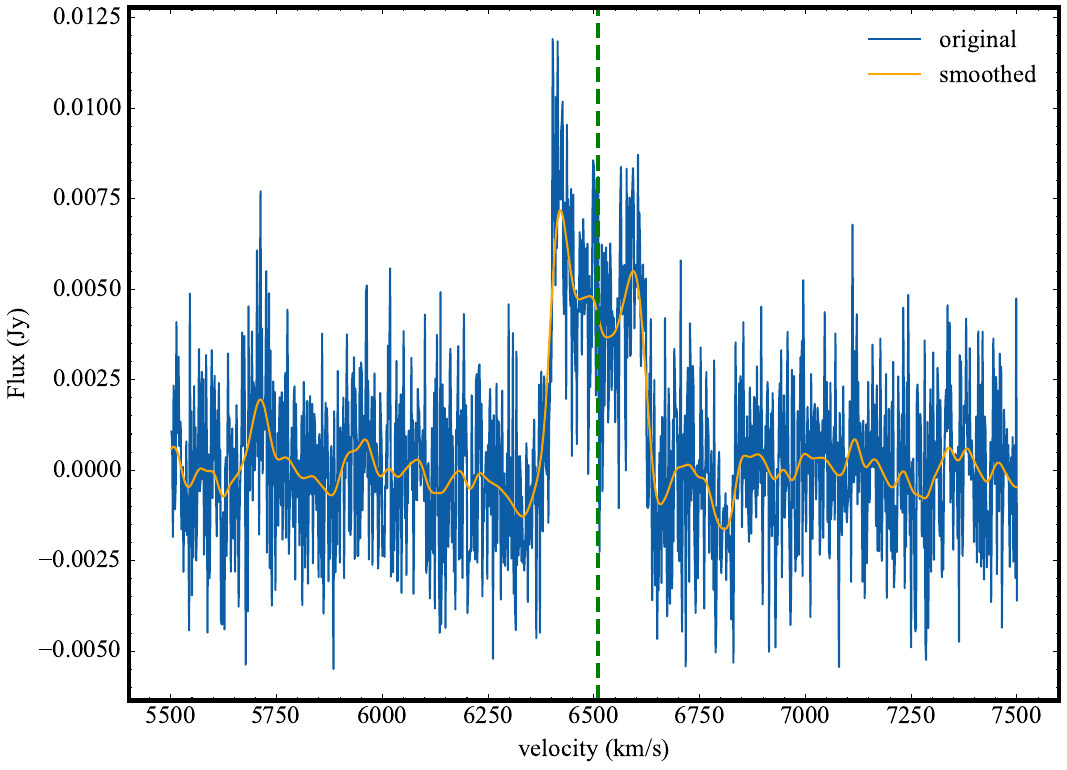}
   \label{9500-6104}
  \end{minipage}
 }
    \subfigure[]
    {
     \begin{minipage}{8cm}
      \centering
      \includegraphics[scale=0.45]{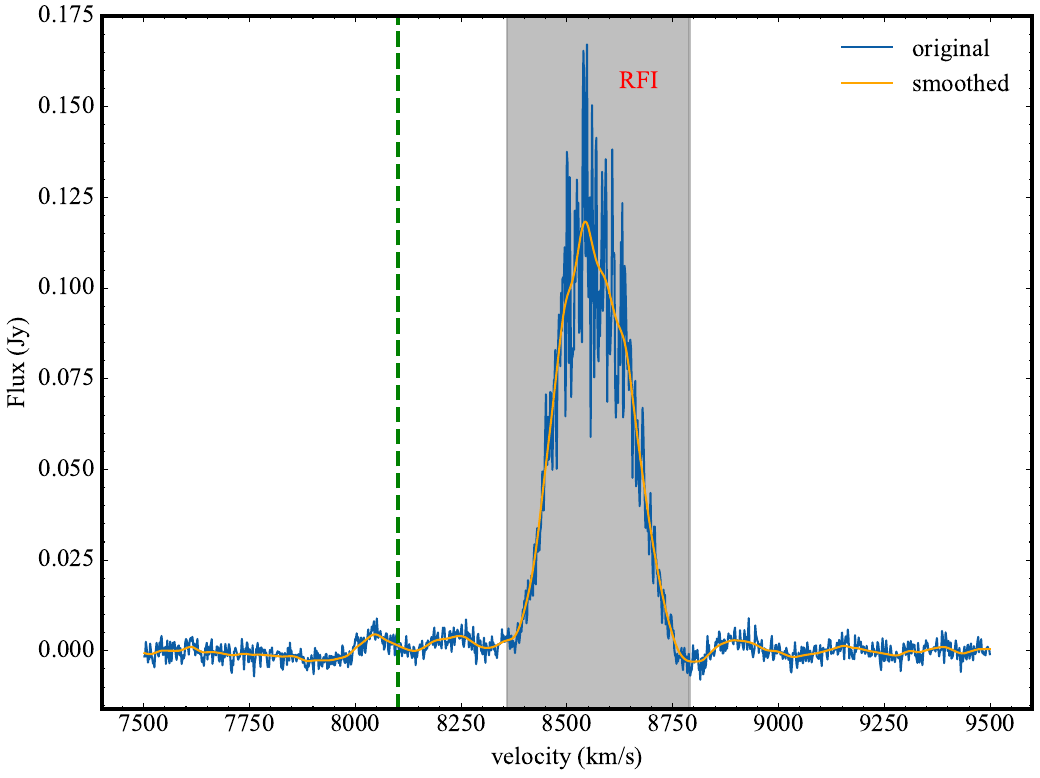}
      \label{9499-9101}
     \end{minipage}
    }
\caption{The example of H\,{\sc i} spectra for our two targets. a) Galaxy PLATEIFU 9500-6104; b) Galaxy PLATEIFU 9499-9101. In this Figure, the blue and orange lines represent the original and smoothed (a resolution of $\sim$ 10 ${\rm km s^{-1}}$) data, respectively. Moreover, the green dashed line is the central velocity of the galaxy (converted by redshift). Radio frequency interference is indicated in the shaded region for the FAST spectra.}
\label{HI example}
\end{figure}

As shown in Fig. \ref{HI example}, we presented examples of H\,{\sc i} spectra for our two targets (PLATEIFU: 9500-6104 and 9499-9101), the latter which is affected by the RFI. In this Figure, the blue and orange lines represent the original and smoothed (a resolution of $\sim$ 10 ${\rm km s^{-1}}$) data, respectively. The green dashed line is the central velocity of the galaxy (converted by redshift). RFI is indicated in the shaded region for the FAST spectra. To confirm the presence of RFI, we not only examined the waterfall plots \citep[see Fig. 2 in][]{2024SCPMA..6759514J} of the H\,{\sc i} spectrum but also checked deeper DESI\footnote{\url{https://viewer.legacysurvey.org/}} images. The waterfall plots display the temperature ($T_{\rm K}$) as a function of frequency and the spectral record number, with RFI typically appearing as bright bands or spots. These signatures can be sudden, periodic, or stable, forming a sharp contrast with the normally smooth and continuous H\,{\sc i} signals \citep[][]{2024SCPMA..6759514J}. RFI may also manifest at specific frequencies or exhibit distinct patterns of variation \citep[][]{2024SCPMA..6759514J}. Our sample information and calculated results for the star-forming S0s are provided in Table \ref{tab1}. From our FAST observations, we found that the S/N of H\,{\sc i} in 11 galaxies out of 41 star-forming S0s is greater than 5, with a detection rate of 27$\%$. 

\subsubsection{Archival Data}

As a supplement, we cross-matched the samples in Table \ref{basic information} with H\,{\sc i}MaNGA catalog \citep[][]{2019MNRAS.488.3396M, 2021MNRAS.503.1345S}. H\,{\sc i}MaNGA is a program of H\,{\sc i} (21cm neutral hydrogen) followup of MaNGA galaxies \citep[redshift $<$ 0.05;][]{2019MNRAS.488.3396M, 2021MNRAS.503.1345S}, and suggests that S/N $>$ 3 indicates the detection of H\,{\sc i} in the galaxy. Consequently, H\,{\sc i}MaNGA covers only a subset of the galaxies (totaling 6623 sources) in MaNGA. In the second column of Table \ref{basic information}, we provide the number of galaxies with redshifts less than 0.05 in each sample. Because our samples are all from MaNGA, so some galaxies don't have H\,{\sc i} information. As shown in Table \ref{basic information}, we found that only 15 sources in star-forming S0s were covered by H\,{\sc i}MaNGA (listed in Table \ref{HIMaNGA}), of which 5 are also part of our selected star-forming S0s (marked with asterisks). However, the S/N of H\,{\sc i} for these 5 sources is less than 5, so we use the calculation results from FAST. For the remaining 10 sources in Table \ref{HIMaNGA}, we used the results provided by H\,{\sc i}MaNGA. 

There are 1381 galaxies with H\,{\sc i} detection among normal blue spirals, but only 923 of these have S/N $>$ 5. In contrast, among 64 $g - r$ red spirals with H\,{\sc i} detection, only 43 sources have S/N $>$ 5. Note that, unless otherwise specified, when we discuss H\,{\sc i} (Sect. \ref{sec3.4}), we focus exclusively on galaxies with H\,{\sc i} detection exhibiting S/N greater than 5; others are treated as upper limits (third column in Table \ref{basic information}). This approach ensures consistency with the observations from FAST (Sect. \ref{2.3.3}).

\begin{deluxetable}{ccccccccccc}
\centering
\tablenum{3}
\small
\tablecaption{15 star-forming S0s in H\,{\sc i}MaNGA\label{HIMaNGA}}
\tablehead{
\begin{tabular}[c]{@{}c@{}}MaNGA$\_$ID\\ (Plate$\_$IFU)\\ (1)\end{tabular} & \begin{tabular}[c]{@{}c@{}}R.A.\\ (deg.)\\ (2)\end{tabular} & \begin{tabular}[c]{@{}c@{}}Decl.\\ (deg.)\\ (3)\end{tabular} & \colhead{\begin{tabular}[c]{@{}c@{}}Redshift\\ ($z$)\\ (4)\end{tabular}} & \begin{tabular}[c]{@{}c@{}}log M$_{\ast}$\\ ($M_{\odot}$)\\ (5)\end{tabular} & \begin{tabular}[c]{@{}c@{}}log SFR$_{\rm SED}$\\ ($M_{\odot}$yr$^{-1}$)\\ (6)\end{tabular} & \begin{tabular}[c]{@{}c@{}}W$_{50}$\\ (km s$^{-1}$)\\ (7)\end{tabular} & \begin{tabular}[c]{@{}c@{}}log M$_{\rm HI}$\\ ($M_{\odot}$)\\ (8)\end{tabular} & \begin{tabular}[c]{@{}c@{}}S/N\\ (9)\end{tabular} & \begin{tabular}[c]{@{}c@{}}rms\\ (mJy)\\ (10)\end{tabular} & \begin{tabular}[c]{@{}c@{}}Obs\\ (11)\end{tabular}
}
\startdata
9183-12705$^{*}$ & 123.10611 & 37.73022 & 0.038 & 10.487 $\pm$ 0.036 & 0.592 $\pm$ 0.08 & 63.74 & 9.57 & 3.84 & 1.70 & GBT\\
8252-3704$^{*}$ & 145.30812 & 47.68860 & 0.047 & 10.526 $\pm$ 0.041 & 0.520 $\pm$ 0.147 & 138.48 & 9.61 & 3.96 & 1.65 & GBT\\
8568-9101$^{*}$ & 155.58076 & 36.58304 & 0.026 & 10.948 $\pm$ 0.031 & 0.527 $\pm$ 0.114 & 381.17 & 9.65 & 4.45 & 1.42 & GBT\\
8441-1901$^{*}$ & 222.86356 & 37.50326 & 0.032 & 10.227 $\pm$ 0.050 & 0.185 $\pm$ 0.207 & 342.01 & 9.75 & 4.47 & 1.39 & GBT\\
8723-6104$^{*}$ & 130.40784 & 54.91850 & 0.045 & 10.640 $\pm$ 0.034 & 0.493 $\pm$ 0.087 & 141.75 & 9.65 & 3.01 & 1.44 & GBT\\
\hline
8625-1901 & 258.43011 & 57.18838 & 0.029 & 9.675 $\pm$ 0.040 & $-0.369$ $\pm$ 0.100 & 292.07 & 9.73 & 4.42 & 1.55 & GBT\\
8615-1902 & 319.75116 & $-0.96399$ & 0.019 & 10.168 $\pm$ 0.026 & $-0.031$ $\pm$ 0.110 & 69.17 & 9.13 & 4.49 & 1.82 & GBT\\
8252-1902 & 146.09184 & 47.45985 & 0.026 & 9.480 $\pm$ 0.045 & $-0.698$ $\pm$ 0.124 & 127.34 & 9.54 & 8.19 & 1.41 & GBT\\
8313-3703 & 241.84668 & 41.70897 & 0.018 & 9.795 $\pm$ 0.028 & $-0.418$ $\pm$ 0.071 & 658.17 & 9.85 & 6.00 & 1.92 & GBT\\
8255-3703 & 166.18780 & 45.15643 & 0.022 & 9.433 $\pm$ 0.032 & $-0.862$ $\pm$ 0.084 & 233.98 & 9.58 & 6.55 & 1.66 & ALFALFA\\
11758-1902 & 203.47165 & 52.70723 & 0.030 & 9.836 $\pm$ 0.049 & $-0.432$ $\pm$ 0.396 & 548.90 & 10.04 & 4.46 & 1.72 & GBT\\
11868-1902 & 250.06174 & 23.64346 & 0.037 & 9.750 $\pm$ 0.056 & $-0.806$ $\pm$ 0.684 & 101.87 & 9.87 & 4.53 & 2.40 & GBT\\
8710-6102 & 117.96618 & 49.81432 & 0.024 & 10.531 $\pm$ 0.031 & $-0.127$ $\pm$ 0.151 & 110.20 & 9.30 & 3.04 & 1.71 & GBT\\
9893-3702 & 256.21393 & 25.05510 & 0.039 & 9.925 $\pm$ 0.050 & $-0.515$ $\pm$ 0.672 & 254.46 & 9.60 & 3.44 & 1.43 & GBT\\
11981-1902 & 254.31749 & 18.45263 & 0.033 & 9.540 $\pm$ 0.024 & $-0.065$ $\pm$ 0.025 & 270.98 & 9.85 & 4.67 & 1.69 & GBT\\
\enddata
\tablecomments{The columns show (1) the MaNGA ID of our targets. The asterisk marks indicate that these galaxies appear simultaneously in Table \ref{tab1}; (2) R.A. in degrees; (3) Decl. in degrees; (4) redshift from NSA; (5), (6) are M$_{\ast}$ and SFR, both from \cite{2016ApJS..227....2S, 2018ApJ...859...11S}; (7), (8), (9), and (10) are estimated H\,{\sc i} spectral parameters \citep[][]{2019MNRAS.488.3396M, 2021MNRAS.503.1345S}; (11) represents the data source.}
\end{deluxetable}

\begin{figure}[!ht]
 \centering
 \subfigure[]
 {
  \begin{minipage}{8cm}
   \centering
   \includegraphics[scale=0.45]{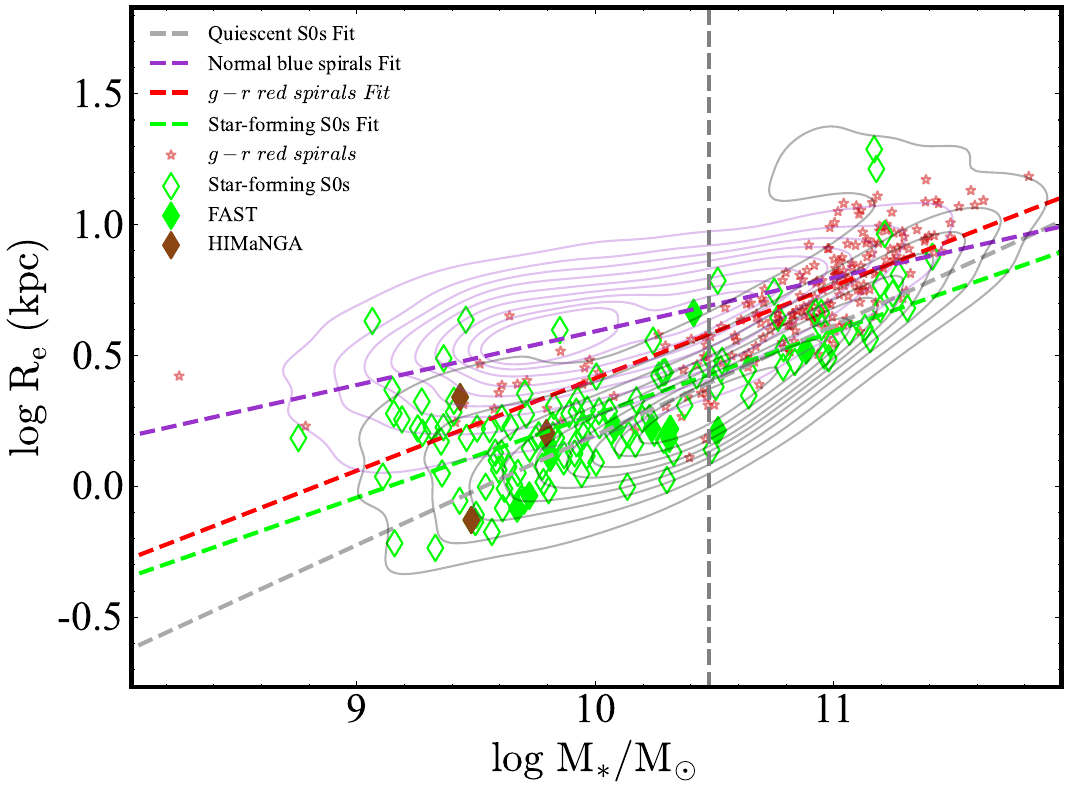}
   \label{size-mass}
  \end{minipage}
 }
    \subfigure[]
    {
     \begin{minipage}{8cm}
      \centering
      \includegraphics[scale=0.45]{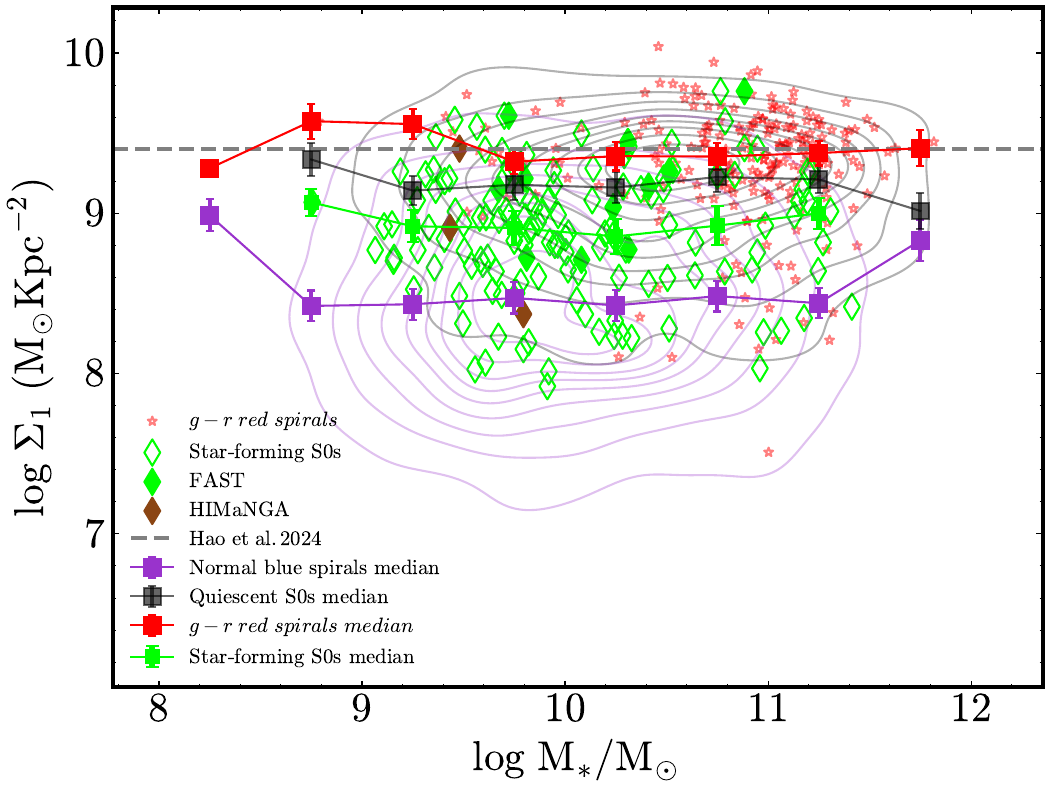}
      \label{sigma1-mass}
     \end{minipage}
    }
\caption{Distributions of our samples in SMR and $\Sigma_{1}$-mass relation. a) SMR. The purple, red, black, and green dashed lines are the fitting results of normal blue spirals, $g - r$ red spirals, quiescent S0s, and star-forming S0s, with slopes of 0.20, 0.35,  0.42, and 0.39, respectively. The black vertical dashed line indicates the critical mass \citep[$3 \times 10^{10} M_{\odot}$;][]{2011MNRAS.413..813C}. b) $\Sigma_{1}$-mass relation. The filled purple squares (normal blue spirals), filled green squares (star-forming S0s), filled black squares (quiescent S0s), and filled red squares ($g - r$ red spirals) represent the median value in the corresponding mass bin respectively, with the error bars showing the error of the median. The black dashed line represents the $\Sigma_{1}$ characteristic value\citep[][]{2024ApJ...968....3H}. In this figure, diamonds and red stars represent star-forming S0s and $g - r$ red spirals respectively, and the filled diamonds represent star-forming S0s of obvious H\,{\sc i} detection (S/N $>$ 5) for FAST (green) and H\,{\sc i}MaNGA (brown), respectively. The purple and black contours are the number density distributions of 2642 normal blue spirals and 680 quiescent S0s, respectively.}
\label{structure properties}
\end{figure}

\section{Results} \label{result}
\subsection{Structural Properties}\label{global-properties}

The size-mass relation (SMR) of galaxies serves as a valuable tool for exploring the formation and evolution of galaxies \citep[e.g.,][]{2013MNRAS.432.1862C, 2014ApJ...788...28V, 2016ARA&A..54..597C, 2024MNRAS.528.2391C, 2024A&A...691A.107C}. \citet{2010ApJ...725.2312O} used large-scale dark matter simulation to recover the observational result of ``archaeological downsizing," indicating that ETGs have a two-phase formation process: in situ SF and external accretion. Of course, earlier studies have also provided empirical evidence prior to this \citep[e.g.,][]{1998MNRAS.297..427A} and a large amount of observational evidence has been obtained to support this claim \citep[e.g.,][]{2011ApJS..195...15D, 2011MNRAS.413.2943F, 2013MNRAS.432.1862C}. Specifically, \citet{2013MNRAS.432.1862C} studied the SMR of 260 ETGs from ATLAS$^{\rm 3D}$ and found that the SMR of galaxies can be described by three characteristic masses (see their Fig. 14), and also found evidence of this ``two-phase" character. \citet{2014ApJ...788...28V} confirmed that ETGs are on average smaller than LTGs at all redshifts using 3D-HST data combined with CANDELS images, and they claimed that galaxies with different SF activities follow different SMR.

We presented the SMR of all our samples in Fig. \ref{size-mass}. In this Figure, the diamonds and open red stars represent star-forming S0s and $g - r$ red spirals, respectively. The least squares fitting results\footnote{We use least squares fitting to trace the correlation between the size and mass of galaxies, known as SMR. During the fitting process, the measurement error of the data is transmitted as weights.\label{foot9}} of $g - r$ red spirals, normal blue spirals, star-forming S0s, and quiescent S0s are given by different color dashed lines, while the black vertical dashed line indicates the characteristic mass \citep[$\sim 3 \times 10^{10} M_{\odot}$;][]{2011MNRAS.413..813C, 2013MNRAS.432.1862C}. In addition, we used the different color contours to represent the number density distributions of 2642 normal blue spirals (purple) and 680 quiescent S0s (black). The filled diamonds with different colors represent star-forming S0s of obvious H\,{\sc i} detection (S/N $>$ 5) for FAST (green) and H\,{\sc i}MaNGA (brown). For $g - r$ red spirals, we do not currently distinguish the S/N of H\,{\sc i} in galaxies. We found that star-forming S0s have steep and warped SMR, which is similar to quiescent S0s. The bending phenomenon of galaxies on SMR is a manifestation of mass quenching, as it occurs near characteristic mass $3 \times 10^{10}M_{\odot}$ \citep[e.g.,][]{2013MNRAS.432.1862C, 2016ARA&A..54..597C, 2024A&A...691A.107C}. This value is closer to the $M_{*}$ threshold, beyond which 50$\%$ of galaxies will be quenched \citep[e.g.,][]{2021ApJ...921...38K, 2024A&A...691A.107C}. The SMR of star-forming S0s mainly covers the regions of quiescent S0s and $g - r$ red spirals, with a significant deviation from the regions of normal blue spirals, and its fitting result is very close to quiescent S0s and $g - r$ red spirals. On average, $g - r$ red spirals are more compact than normal blue spirals and have a steeper SMR, similar to quiescent S0s and star-forming S0s. Furthermore, $g - r$ red spirals also exhibit bending at the same mass ($\sim 3 \times 10^{10} M_{\odot}$).

How galaxies quench is a key issue in understanding galaxy evolution, and both theoretical and observational studies indicate that the formation of a dense core at the center of the galaxy is a necessary condition for galaxy quenching \citep[e.g.,][]{2013ApJ...776...63F, 2017ApJ...840...47B, 2020ApJ...897..162G, 2024ApJ...968....3H}. The stellar mass surface density within 1 kpc ($\Sigma_{1}$), which is widely used as a tracer for the dense core at the center of galaxies, is a key parameter connecting the history of galaxy formation \citep[e.g.,][]{2013ApJ...776...63F, 2017ApJ...840...47B, 2020ApJ...897..162G, 2024ApJ...968....3H}. \citet{2013ApJ...776...63F} found that SFGs have steeper $\Sigma_{1}$-$M_{\ast}$ relation than GVGs/red-sequence galaxies using data from SDSS DR7 and $Galaxy\ Evolution\ Explorer$ ($GALEX$) survey. \citet{2020ApJ...897..162G} also found similar result, that galaxies with high $\Sigma_{1}$ are expected to be quenched.

\begin{figure}[!ht]
 \centering
 \subfigure[]
 {
  \begin{minipage}{8cm}
   \centering
   \includegraphics[scale=0.45]{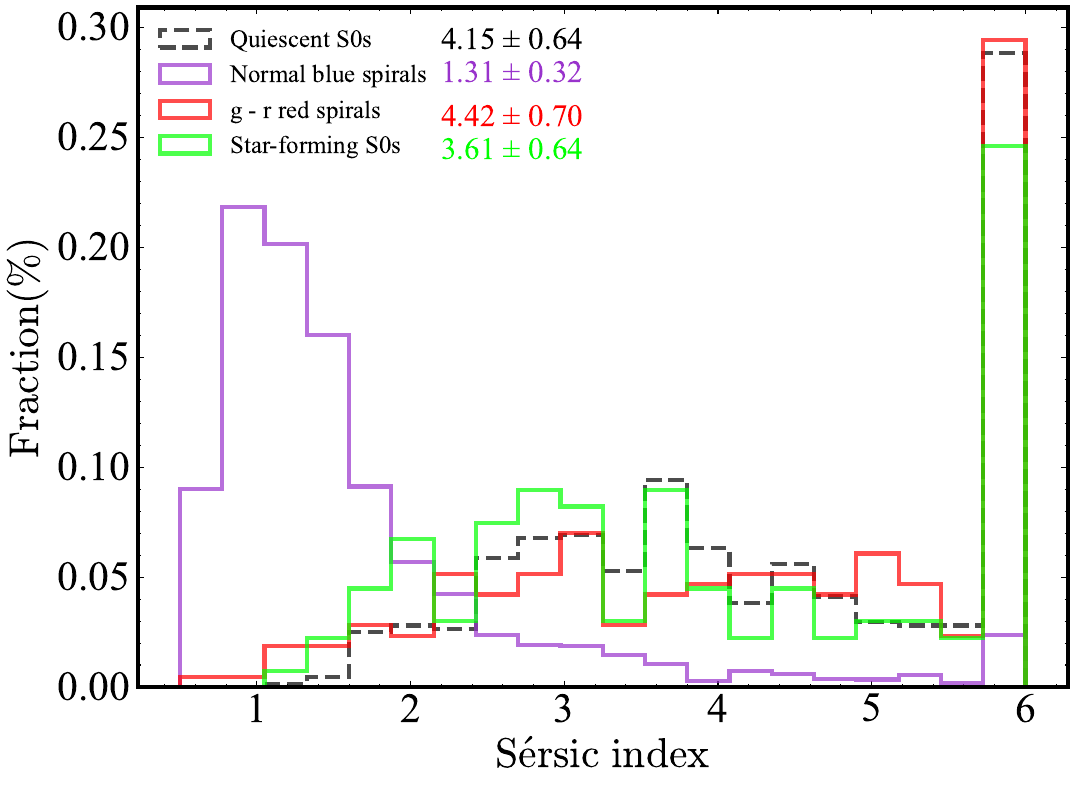}
   \label{sersic-index}
  \end{minipage}
 }
    \subfigure[]
    {
     \begin{minipage}{8cm}
      \centering
      \includegraphics[scale=0.45]{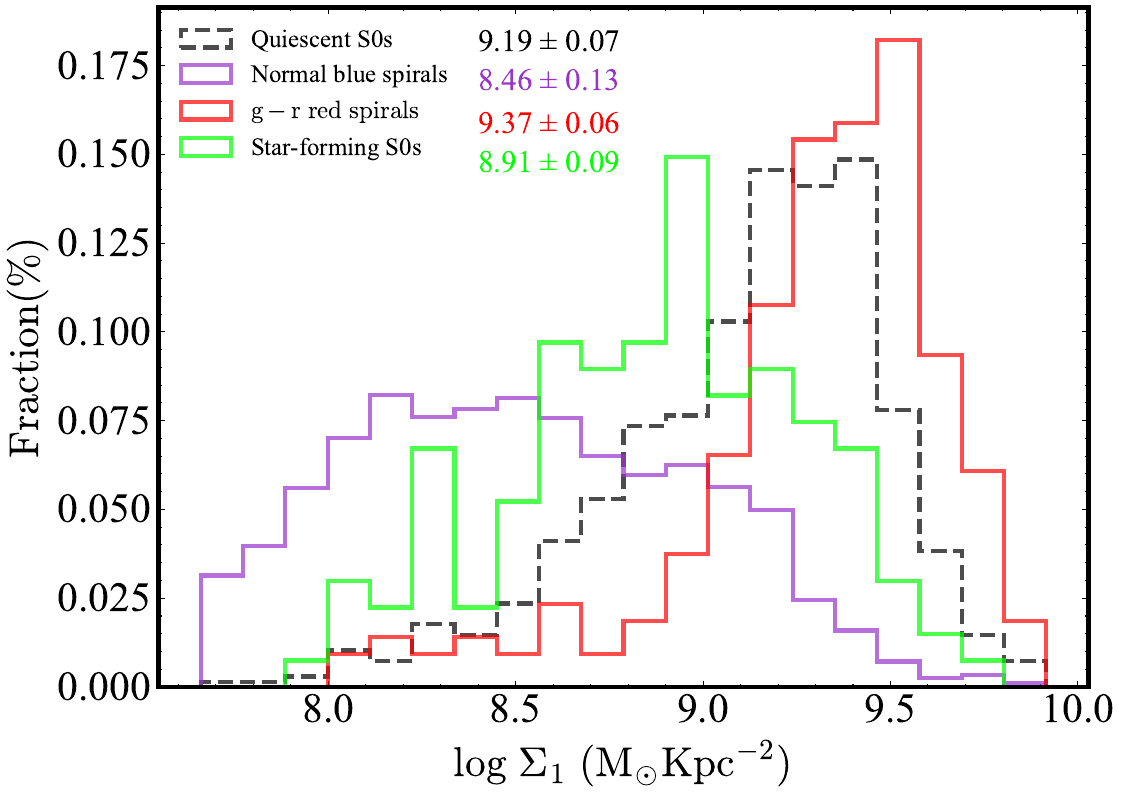}
      \label{sigma1-hist}
     \end{minipage}
    }
\caption{Histograms of S$\acute{\rm e}$rsic index and $\Sigma_{1}$. a) S$\acute{\rm e}$rsic index. b) $\Sigma_{1}$. In this Figure, the different color numbers are the median values. The colors and labels are the same as in Fig. \ref{structure properties}.}
\label{sersic-sigma1}
\end{figure}

In Fig. \ref{sigma1-mass}, we provided the $\Sigma_{1}$-$M_{\ast}$ relation. The open marks represent the distribution of samples, and the filled marks represent the median value in the corresponding mass bin. The black dashed line represents the $\Sigma_{1}$ characteristic value ($\Sigma_{1}$ = $10^{9.4}$ $M_{\odot}kpc^{-2}$), used to distinguish blue spirals with high and low $\Sigma_{1}$ \citep[][]{2024ApJ...968....3H}. The other labels are the same as in Fig. \ref{size-mass}. We found that the distribution of $g - r$ red spirals in the $\Sigma_{1}$-$M_{\ast}$ relation is close to that of star-forming/quiescent S0s. On average, $g - r$ red spirals, quiescent S0s, and star-forming S0s have higher $\Sigma_{1}$ than normal blue spirals. On both low- and high-mass ends, the significant increase in $\Sigma_{1}$ of normal blue spirals is due to the insufficient number of galaxies. There are few galaxies ($\sim$ 2.35$\%$) in normal blue spirals with $\Sigma_{1}$ exceeding the characteristic value. The Kolmogorov-Smirnov (KS) test shows that there are significant statistical differences in the distribution of $\Sigma_{1}$ among our four samples ( see Table 4), but the proportion of dense cores in $g - r$ red spirals and quiescent S0s is significantly higher (see Table \ref{basic information}). Moreover, the Anderson-Darling \citep[as detailed in][]{ADTest} test (AD test) and the permutation test \citep[][]{rizzo2019statistical} also show the same results (see Table \ref{test_result}). The S$\acute{\rm e}$rsic index (Fig. \ref{sersic-index}) of normal blue spirals is mostly (79$\pm$5$\%$) less than 2 (see Table \ref{basic information}). The KS test shows that there is no significant statistical difference in the distribution of S$\acute{\rm e}$rsic index between $g - r$ red spirals and quiescent S0s (p-value: 0.37), while the statistical differences between the distributions of the other samples are significant (Table \ref{test_result}\footnote{We provided the results of KS test, AD test, and Permutation test for the statistical tests of all variables in this paper.}). For comparison, the histogram of $\Sigma_{1}$ is shown in Fig. \ref{sigma1-hist}, and the corresponding color numbers give the sample median values.

\begin{deluxetable}{cccccccc}
\centering
\tablenum{4}
\tiny
\tablecaption{Statistical tests for different samples\label{test_result}}
\tablehead{
\begin{tabular}[c]{@{}c@{}}Parameters\\ (1)\end{tabular} & \begin{tabular}[c]{@{}c@{}}Normal blue spirals\\ (median)\\ (2)\end{tabular} & \begin{tabular}[c]{@{}c@{}}$g - r$ red spirals\\ (median)\\ (3)\end{tabular} & \colhead{\begin{tabular}[c]{@{}c@{}}Star-forming S0s\\ (median)\\ (4)\end{tabular}} & \begin{tabular}[c]{@{}c@{}}Quiescent S0s\\ (median)\\ (5)\end{tabular} & \begin{tabular}[c]{@{}c@{}}KS test\\ (p-value)\\ (6)\end{tabular} & \begin{tabular}[c]{@{}c@{}}AD test\\ (significance level)\\ (7)\end{tabular} & \begin{tabular}[c]{@{}c@{}}Permutation test\\ (pvalue)\\ (8)\end{tabular}
}
\startdata
S$\acute{\rm e}$rsic index & 1.31$\pm$0.32 & 4.42$\pm$0.70 & 3.61$\pm$0.64 & 4.15$\pm$0.64 & \begin{tabular}[c]{@{}c@{}}8.54$\times$10$^{-32}$$_{(2, 3)}$\\ 5.26$\times$10$^{-27}$$_{(2, 4)}$\\ 1.03$\times$10$^{-21}$$_{(2, 5)}$\\ 0.02$_{(3, 4)}$\\
0.37$_{(3, 5)}$\\
0.02$_{(4, 5)}$\end{tabular} & \begin{tabular}[c]{@{}c@{}}0.001$_{(2, 3)}$\\ 0.001$_{(2, 4)}$\\ 0.001$_{(2, 5)}$\\ 0.03$_{(3, 4)}$\\
0.12$_{(3, 5)}$\\
0.03$_{(4, 5)}$\end{tabular} & \begin{tabular}[c]{@{}c@{}} $<$0.01$_{(2, 3)}$\\ $<$0.01$_{(2, 4)}$\\ $<$0.01$_{(2, 5)}$\\ 0.04$_{(3, 4)}$\\
0.82$_{(3, 5)}$\\
0.009$_{(4, 5)}$\end{tabular}\\
\hline
$\Sigma_{1}$ & 8.46$\pm$0.13 & 9.37$\pm$0.0.06 & 8.91$\pm$0.09 & 9.19$\pm$0.07 & \begin{tabular}[c]{@{}c@{}}3.54$\times$10$^{-27}$$_{(2, 3)}$\\ 5.64$\times$10$^{-18}$$_{(2, 4)}$\\ 3.54$\times$10$^{-21}$$_{(2, 5)}$\\ 1.05$\times$10$^{-22}$$_{(3, 4)}$\\
3.57$\times$10$^{-11}$$_{(3, 5)}$\\
1.46$\times$10$^{-10}$$_{(4, 5)}$\end{tabular} & \begin{tabular}[c]{@{}c@{}}0.001$_{(2, 3)}$\\ 0.001$_{(2, 4)}$\\ 0.001$_{(2, 5)}$\\ 0.001$_{(3, 4)}$\\
0.001$_{(3, 5)}$\\
0.001$_{(4, 5)}$\end{tabular} & \begin{tabular}[c]{@{}c@{}} $<$0.01$_{(2, 3)}$\\ $<$0.01$_{(2, 4)}$\\ $<$0.01$_{(2, 5)}$\\ $<$0.01$_{(3, 4)}$\\
$<$0.01$_{(3, 5)}$\\
$<$0.01$_{(4, 5)}$\end{tabular}\\
\hline
$M_{\rm halo}$ & 12.20$\pm$0.10 & 12.74$\pm$0.12 & 12.73$\pm$0.16 & 12.66$\pm$0.16 & \begin{tabular}[c]{@{}c@{}}1.79$\times$10$^{-18}$$_{(2, 3)}$\\ 9.90$\times$10$^{-7}$$_{(2, 4)}$\\ 1.39$\times$10$^{-19}$$_{(2, 5)}$\\ 0.46$_{(3, 4)}$\\
0.09$_{(3, 5)}$\\
0.99$_{(4, 5)}$\end{tabular} & \begin{tabular}[c]{@{}c@{}}0.001$_{(2, 3)}$\\ 0.001$_{(2, 4)}$\\ 0.001$_{(2, 5)}$\\ 0.22$_{(3, 4)}$\\
0.06$_{(3, 5)}$\\
0.25$_{(4, 5)}$\end{tabular} & \begin{tabular}[c]{@{}c@{}}$<$0.01$_{(2, 3)}$\\ $<$0.01$_{(2, 4)}$\\ $<$0.01$_{(2, 5)}$\\ 0.71$_{(3, 4)}$\\
0.29$_{(3, 5)}$\\
0.76$_{(4, 5)}$\end{tabular}\\
\hline
$M_{\rm HI}$ & 9.67$\pm$0.08 & 9.80$\pm$0.07 & 9.65$\pm$0.07 & -- & \begin{tabular}[c]{@{}c@{}}0.004$_{(2, 3)}$\\ 0.24$_{(2, 4)}$\\ --\\ 0.03$_{(3, 4)}$\\
--\\
--\end{tabular} & \begin{tabular}[c]{@{}c@{}}0.005$_{(2, 3)}$\\ 0.25$_{(2, 4)}$\\ --\\ 0.003$_{(3, 4)}$\\
--\\
--\end{tabular} & \begin{tabular}[c]{@{}c@{}} $<$0.01$_{(2, 3)}$\\ 0.56$_{(2, 4)}$\\ --\\ 0.03$_{(3, 4)}$\\
--\\
--\end{tabular}\\
\enddata
\tablecomments{In this table, we provide the median values of different parameters of samples, as well as the results of statistical tests (KS test; AD test; Permutation test). The number in parenthesis indicates the corresponding column of samples used for the two-sample test. For example, (2, 3) represents a statistical test between normal blue spirals and $g - r$ red spirals. If the significance level of the AD test \citep[][]{ADTest} is 0.001, it indicates that the statistical difference between the two variables is very significant \citep[e.g.,][]{2024ApJ...961...93S}. The column 8 is the permutation test \citep[][]{rizzo2019statistical}. If the value of the Permutation test is less than 0.01, it indicates that the L1 energy distance between samples is large, i.e., the statistical difference is very significant \citep[][]{rizzo2019statistical}.}
\end{deluxetable}

\subsection{Stellar Populations Properties}\label{SPs properties}

We examined the radial profiles of the spectral indices ($D_{\rm n}$4000 and [Mgb/Fe]) for the galaxy. $D_{\rm n}$4000 (the ratio of the average flux density in the bands 4000-4100\,\AA\ and 3850-3950\,\AA\ ) is a proxy for the galaxy age, strongly dependent on metallicity \citep[e.g.,][]{2003MNRAS.344.1000B}. In addition, the $\alpha$-elements are mainly from Type II supernova explosions of massive stars, while a substantial fraction of Fe-peak elements come from the delayed Type I supernova explosions \citep[][]{2005ApJ...621..673T}. Therefore, their ratio ([Mgb/Fe] = Mgb/0.5(Fe5270 + Fe5335)) can reflect the importance of two types of supernovae in the galaxy, carrying the SF timescale in the galaxy \citep[e.g.,][]{2005ApJ...621..673T, 2020ApJ...897..162G, 2019ApJ...883L..36H, 2023ApJ...948...96C, 2024MNRAS.528.2391C}. Usually, the longer the SF timescale of a galaxy, the lower this ratio \citep[][]{2005ApJ...621..673T}.

\begin{figure}[ht]
 \centering
 \subfigure[]
 {
  \begin{minipage}{8cm}
   \centering
   \includegraphics[scale=0.45]{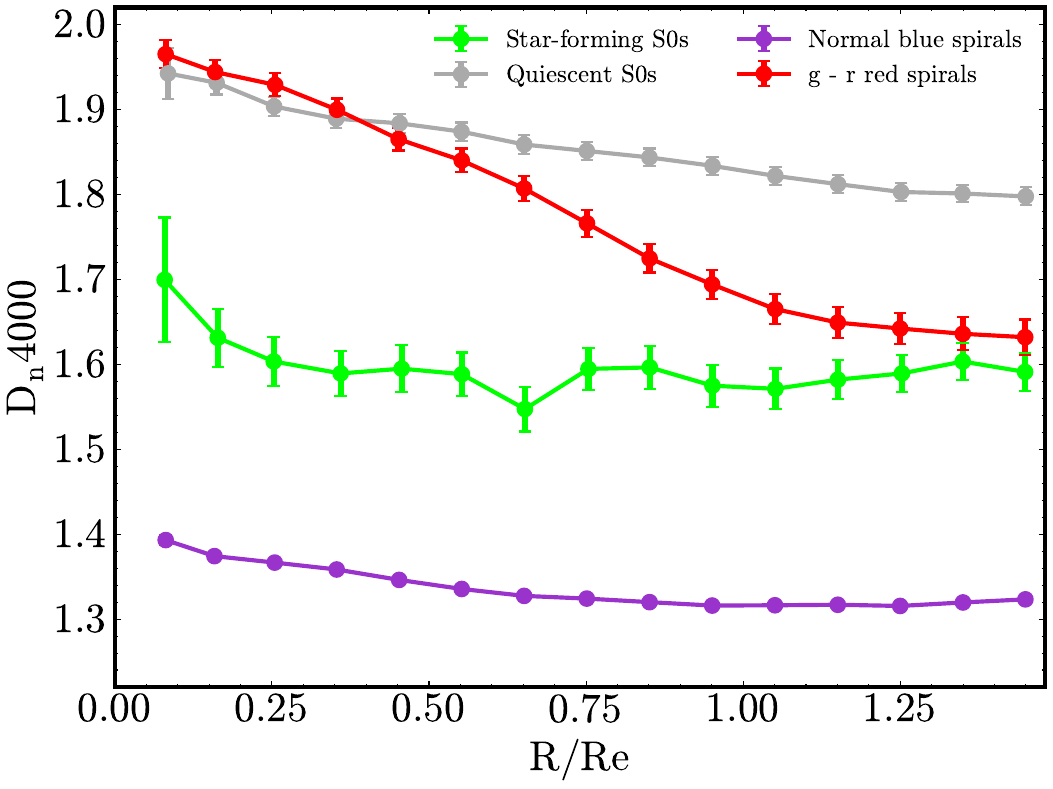}
   \label{dn4000-rp}
  \end{minipage}
 }
    \subfigure[]
    {
     \begin{minipage}{8cm}
      \centering
      \includegraphics[scale=0.45]{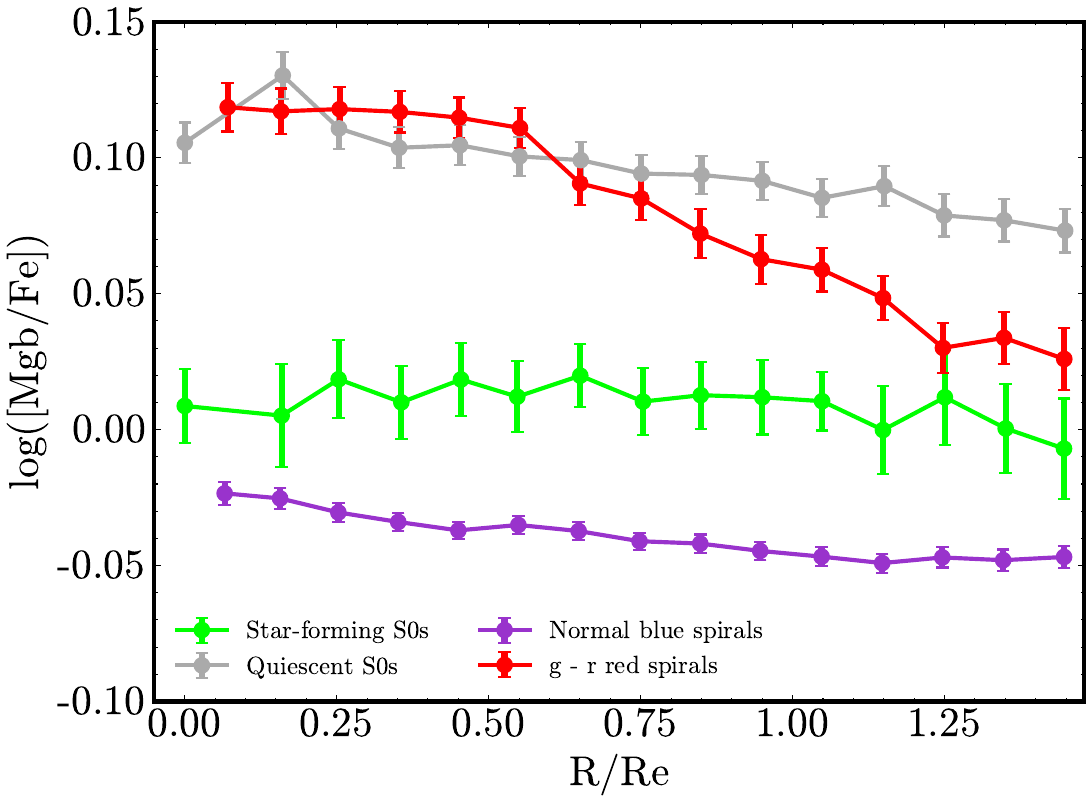}
      \label{mgbfe-rp}
     \end{minipage}
    }
\caption{Medain radial profiles of $D_{\rm n}4000$ and [Mgb/Fe]. a) $D_{\rm n}4000$. b) [Mgb/Fe]. In this Figure, the value of each galaxy in each radial bin is determined as the median value of the pixels within that bin. Obtain the median radial profile by calculating the median of all galaxies of each type. The colors and labels are the same as in Fig. \ref{structure properties}, and the error bars show the error of the median.}
\label{dn4000-mgbfe-rp}
\end{figure}

The radial profiles of $D_{\rm n}$4000 and [Mgb/Fe] are shown in Fig. \ref{dn4000-mgbfe-rp}. The color and labels are the same as in Figure \ref{structure properties}. In Fig. \ref{dn4000-rp}, we found that quiescent S0s and $g - r$ red spirals both exhibit similar old centers, but $g - r$ red spirals have a steeper profile, which is the same as \citet{2024MNRAS.528.2391C}. Both star-forming S0s and normal blue spirals exhibit relatively flat $D_{\rm n}$4000 gradients. Furthermore, we found that $g - r$ red spirals and quiescent S0s have similar higher [Mgb/Fe] ratios in their central regions, but [Mgb/Fe] decreases in the outer regions of $g - r$ red spirals in Fig. \ref{mgbfe-rp}. Similarly, the [Mgb/Fe] radial profiles of star-forming S0s and normal blue spirals remain flat. 

\subsection{Galaxy Halo Mass}

Many studies have shown that the environments of galaxies play a key role in the formation and evolution of galaxies \citep[e.g.,][]{1980ApJ...236..351D, 2006PASP..118..517B, 2013pss6.book..207V}. \citet{2014MNRAS.440..889S} found that LTGs in massive halos (${\rm log}M_{\rm halo} > 12\ M_{\odot}/h$) are mainly located in the red sequence, while galaxies in less massive halos form the blue cloud. \citet{2020ApJ...897..162G} found that over 80$\%$ of $NUV - r$ selected massive red spirals (log$M_{\ast}$ $>$ 10.5$M_{\odot}$) have halo mass greater than the critical halo mass of $10^{12} M_{\odot}/h$. Moreover, \citet{2010MNRAS.405..783M} found that red spirals preferentially in intermediate density regimes.

\begin{figure}[!ht]
 \centering
     \subfigure[]
    {
     \begin{minipage}{8cm}
      \centering
      \includegraphics[scale=0.45]{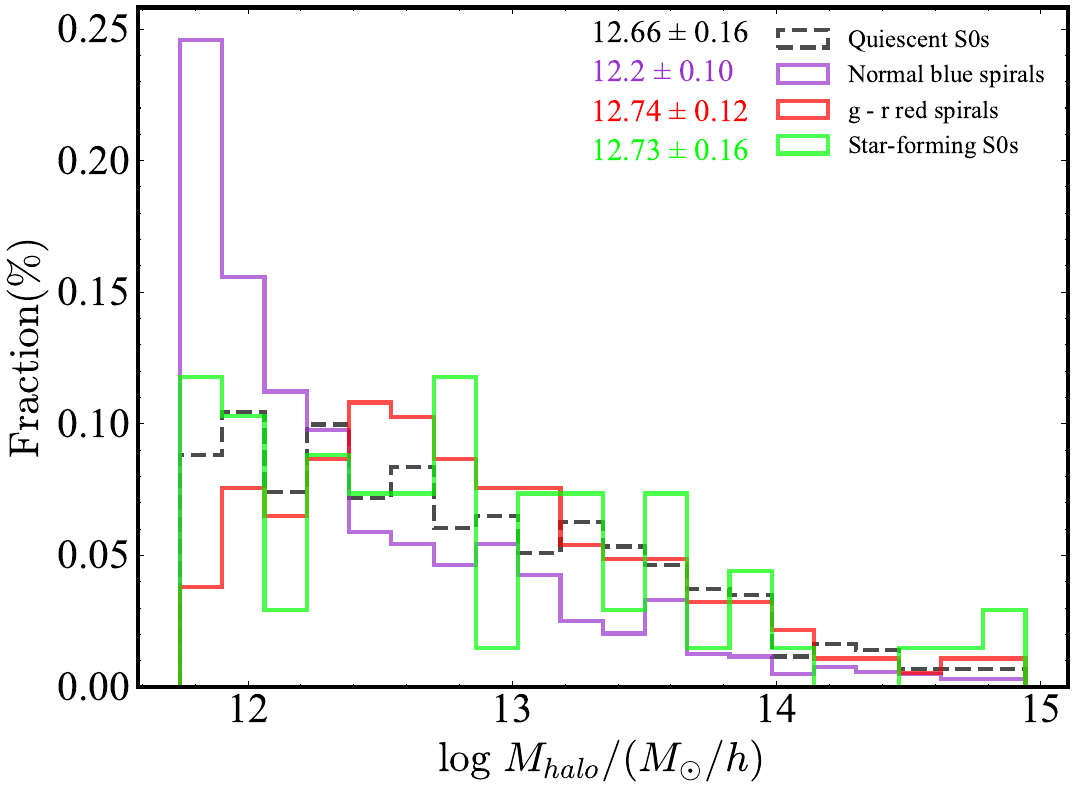}
      \label{Halo-hist}
     \end{minipage}
    }
    \subfigure[]
    {
     \begin{minipage}{8cm}
      \centering
      \includegraphics[scale=0.45]{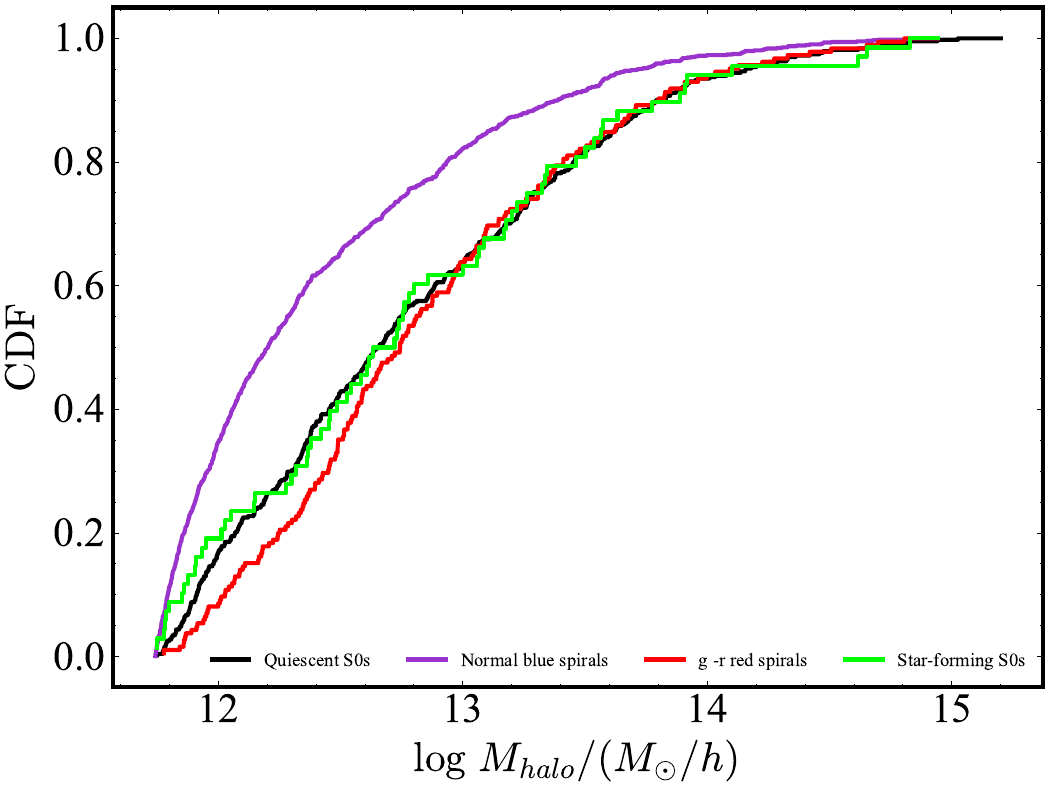}
      \label{CDF_cumulative}
     \end{minipage}
    }
\caption{Histogram of galaxies halo mass and its cumulative distribution function (CDF). a) Histogram of the galaxy halo mass. The different color numbers represent the median values of halo mass. b) CDF. In this figure, the colors and labels are the same as in Fig. \ref{structure properties}. The halo mass is from \citet{2007ApJ...671..153Y}.}
\label{Halo_mass}
\end{figure}

Fig. \ref{Halo_mass} shows the histogram and cumulative distribution function (CDF) of halo mass for our samples, according to the catalog of \citet{2007ApJ...671..153Y}. Their method of deriving halo mass is based on the halo occupation model \citep[e.g.,][]{1998ApJ...494....1J, 2003MNRAS.339.1057Y}, which utilizes the observed galaxy luminosity function and two-point correlation functions to constrain the average number of galaxies of given properties that occupy a dark matter halo of given mass. The color and labels are the same as in Fig. \ref{structure properties}. There are 64$\pm$6$\%$ normal blue spirals, 66$\pm$5$\%$ quiescent S0s, 54$\pm$5$\%$ star-forming S0s, and 77$\pm$8$\%$ $g - r$ red spirals central galaxies, which are labeled as brightest in \citet{2007ApJ...671..153Y}. This is similar to the previous \citep[e.g.,][]{2024MNRAS.528.2391C}. We found that over 82$\pm$5$\%$ of galaxies in star-forming S0s have halo mass exceeding the critical mass, and there is no significant statistical difference between their halo mass and that of quiescent S0s (p-value: 0.99) and $g - r$ red spirals (p-value: 0.46). In addition, quiescent S0s and $g - r$ red spirals also show no significant difference (p-value: 0.09), and the halo mass of most quiescent S0s (83$\pm$5$\%$) and red spirals (92$\pm$7$\%$) exceeds the critical mass. The statistical differences between the distribution of these three and normal blue spirals are significant (p-value $<$ $1.0 \times 10^{-7}$, Table \ref{test_result}).

\begin{figure}[!ht]
 \centering
     \subfigure[]
    {
     \begin{minipage}{8cm}
      \centering
      \includegraphics[scale=0.45]{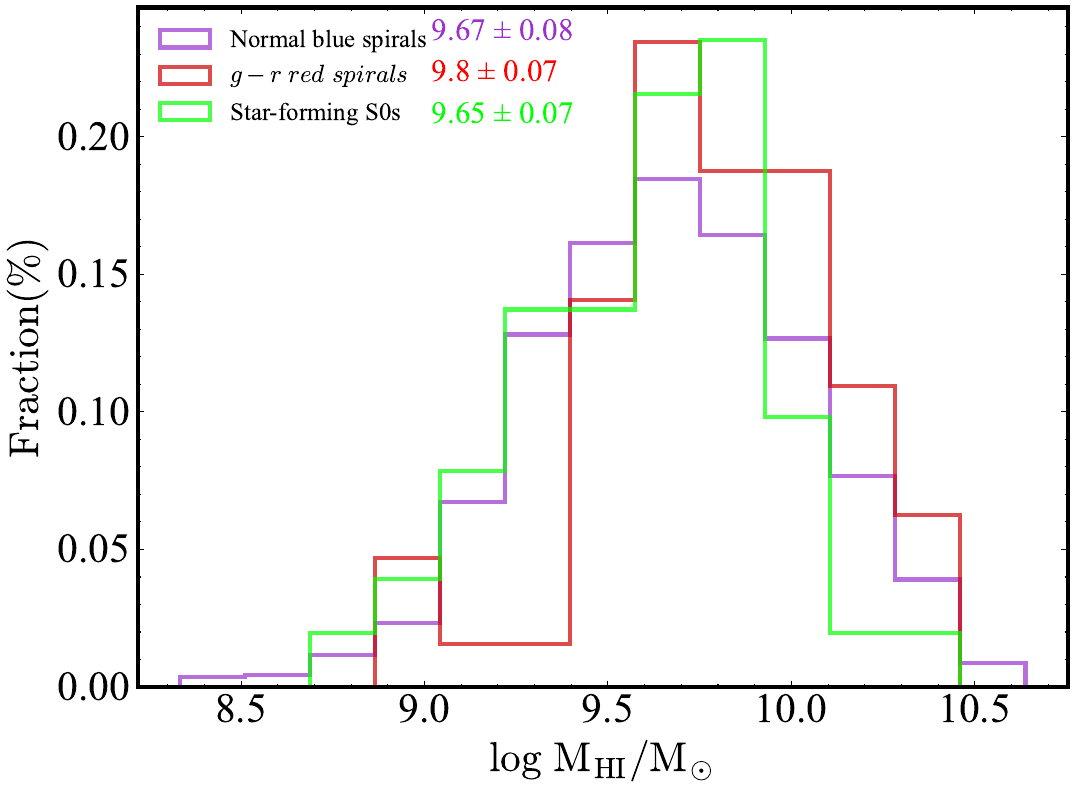}
      \label{MHI-hist}
     \end{minipage}
    }
    \subfigure[]
    {
     \begin{minipage}{8cm}
      \centering
      \includegraphics[scale=0.45]{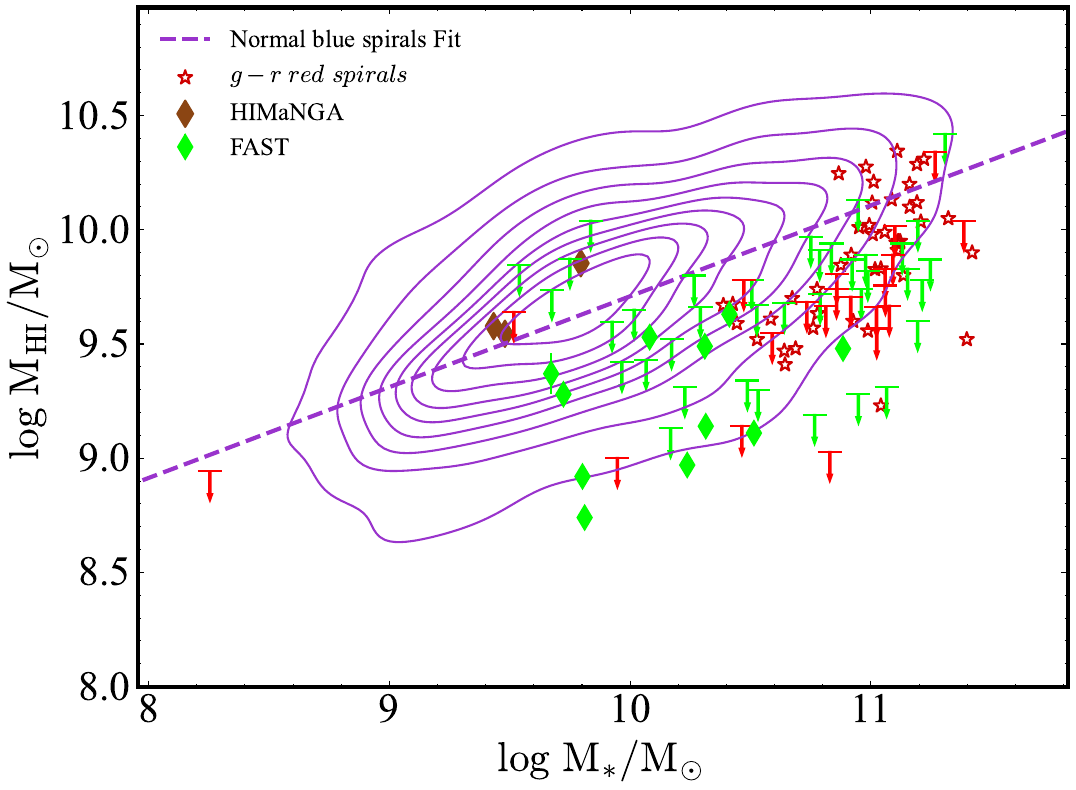}
      \label{MHI-mass}
     \end{minipage}
    }
\caption{Histogram of H\,{\sc i} mass ($M_{\rm HI}$) and the correlation between $M_{\rm HI}$ and $M_{\ast}$. a) The histogram of $M_{\rm HI}$. The numbers with different colors represent the median value. b) $M_{\rm HI}$-$M_{\ast}$. The downward arrows are the upper limit. The purple dashed line represents the fitting of normal blue spirals, with a slope of 0.45 $\pm$ 0.03 and an intercept of 5.26 $\pm$ 0.09. In this figure, the colors and labels are the same as in Fig. \ref{structure properties}. The purple contours represent the number density distribution of 923 normal blue spirals (S/N of H\,{\sc i} $>$ 5).}
\label{MHI-hist-mass}
\end{figure}

\subsection{Neutral Hydrogen Content}\label{sec3.4}

Studies have shown that galaxies are formed through the cooling and condensation of gas at the center of dark matter halos \citep[e.g.,][]{2024ApJ...963...86L}. While molecular neutral hydrogen (H$_{2}$) is considered a raw material for SF, atomic neutral hydrogen (H\,{\sc i}) can be converted into H$_{2}$ \citep[e.g.,][]{2008AJ....136.2846B, 2010AJ....140.1194B}. Furthermore, most of the cold gas in the universe is in the form of H\,{\sc i}, making the reservoir of H\,{\sc i} crucial for regulating the rise and fall of SF. In addition, H\,{\sc i} can be easily observed through the 21 cm hyperfine emission line. Therefore, exploring the reservoir of H\,{\sc i} through large 21 cm surveys is very interesting, e.g., the Arecibo Fast Legacy ALFA Survey \citep[][]{2018ApJ...861...49H}, and the GALEX Arecibo SDSS Survey \citep[GASS;][]{2018MNRAS.476..875C}. Currently, extensive surveys involving multiband imaging and spectroscopy, along with H\,{\sc i} emission at 21 cm, have revealed a strong correlation between H\,{\sc i} mass and the optical/UV properties of galaxies \citep[e.g.,][]{2012ApJ...756..113H, 2015MNRAS.452.2479B, 2018MNRAS.476..875C, 2022ApJ...941...48L}. It is widely believed that the cessation of SF in galaxies is closely linked to a decrease in the H\,{\sc i} reservoir.

\begin{figure}[!htbp]
 \centering
    \subfigure[]
    {
     \begin{minipage}{8cm}
      \centering
      \includegraphics[scale=0.45]{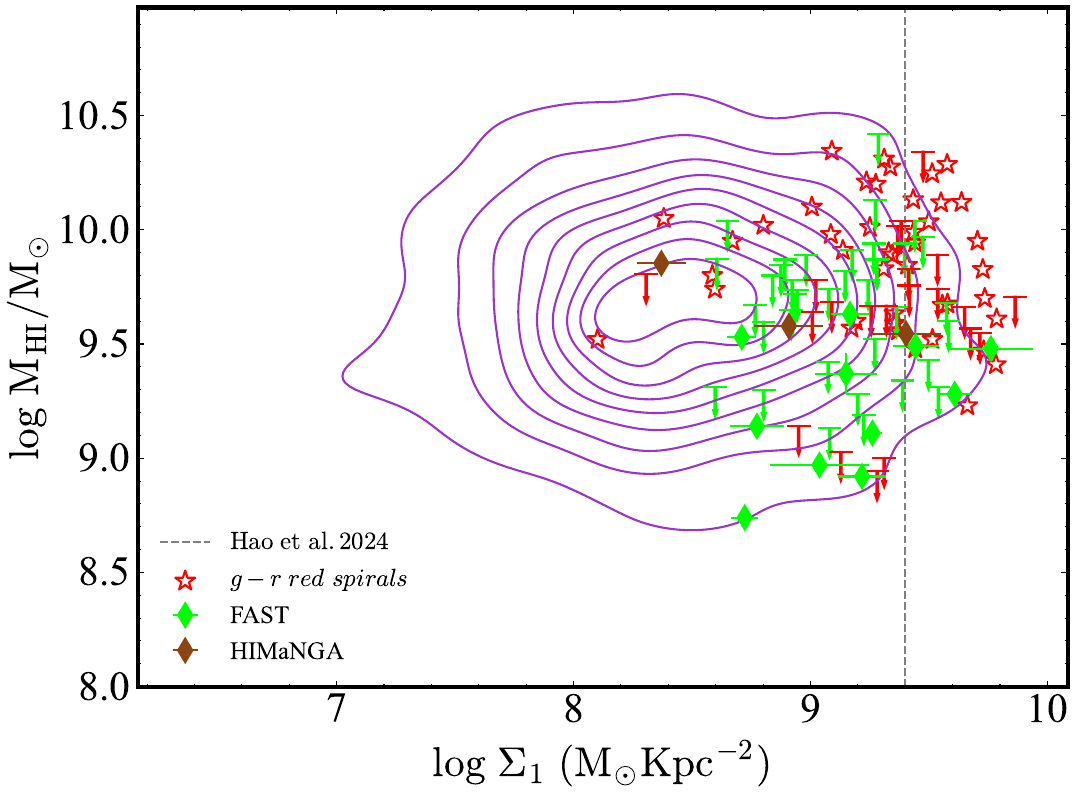}
      \label{MHI-sigma1}
     \end{minipage}
    }
    \subfigure[]
    {
     \begin{minipage}{8cm}
      \centering
      \includegraphics[scale=0.45]{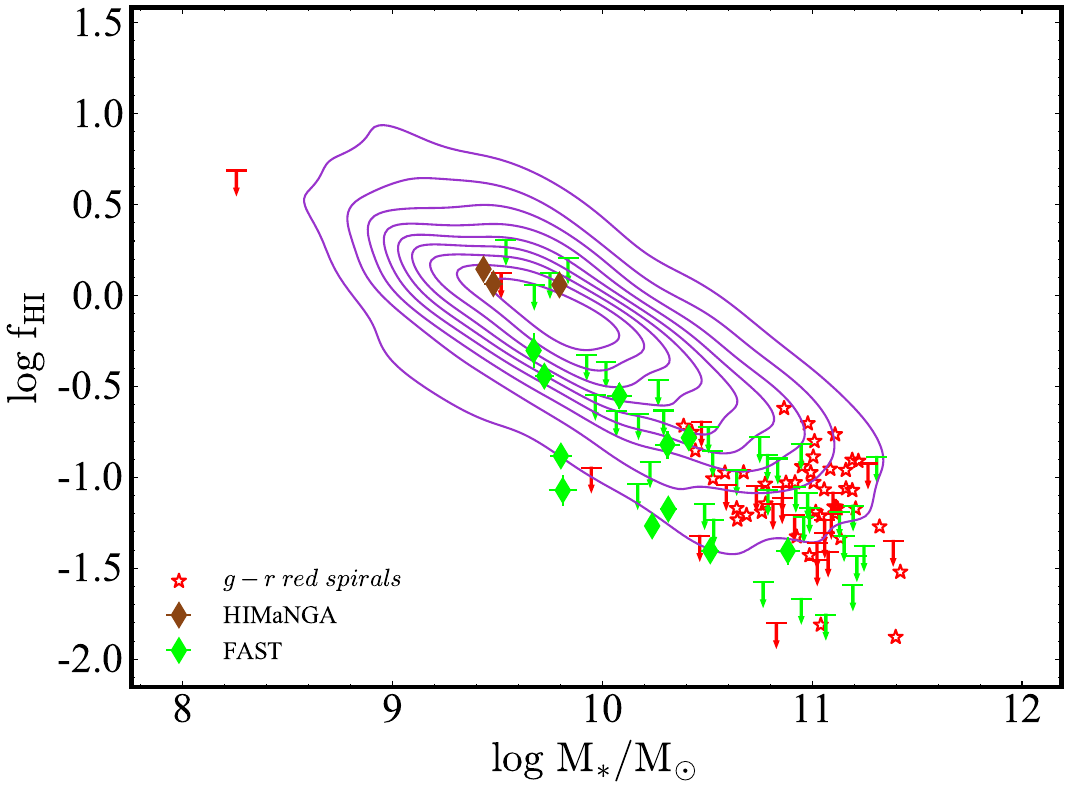}
      \label{fHI-Mass}
     \end{minipage}
    }
    \subfigure[]
    {
     \begin{minipage}{8cm}
      \centering
      \includegraphics[scale=0.45]{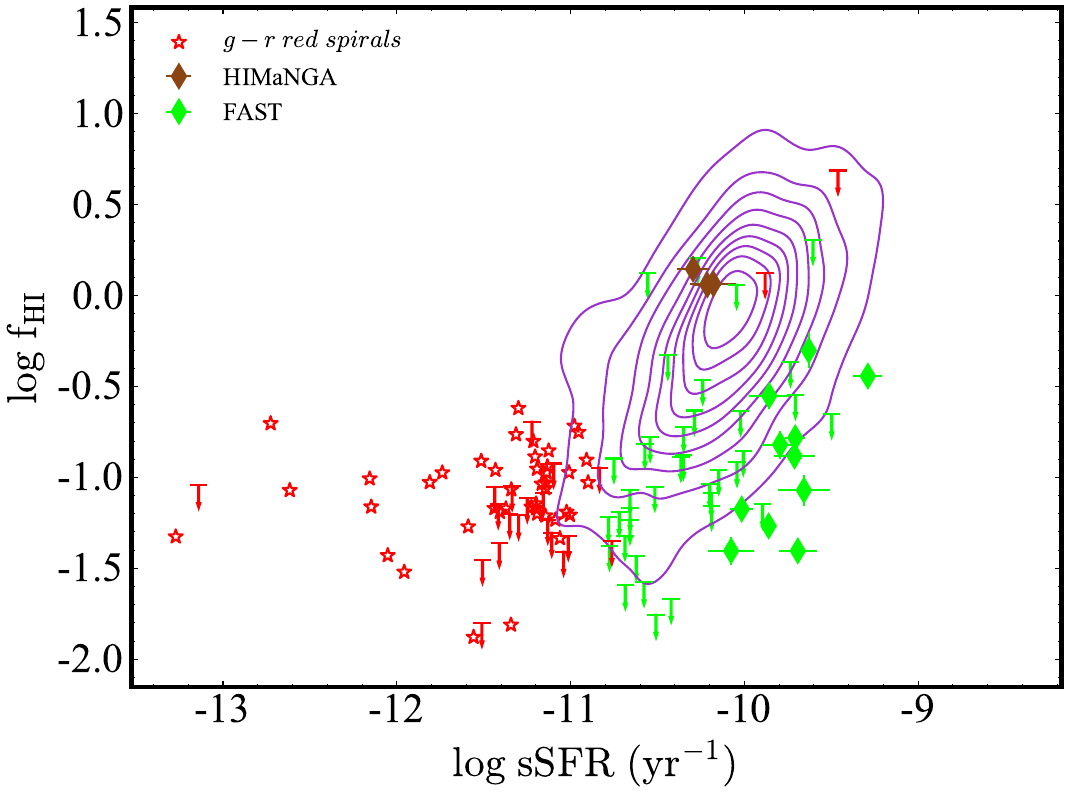}
      \label{fHI-sSFR}
     \end{minipage}
    }
    \subfigure[]
    {
     \begin{minipage}{8cm}
      \centering
      \includegraphics[scale=0.45]{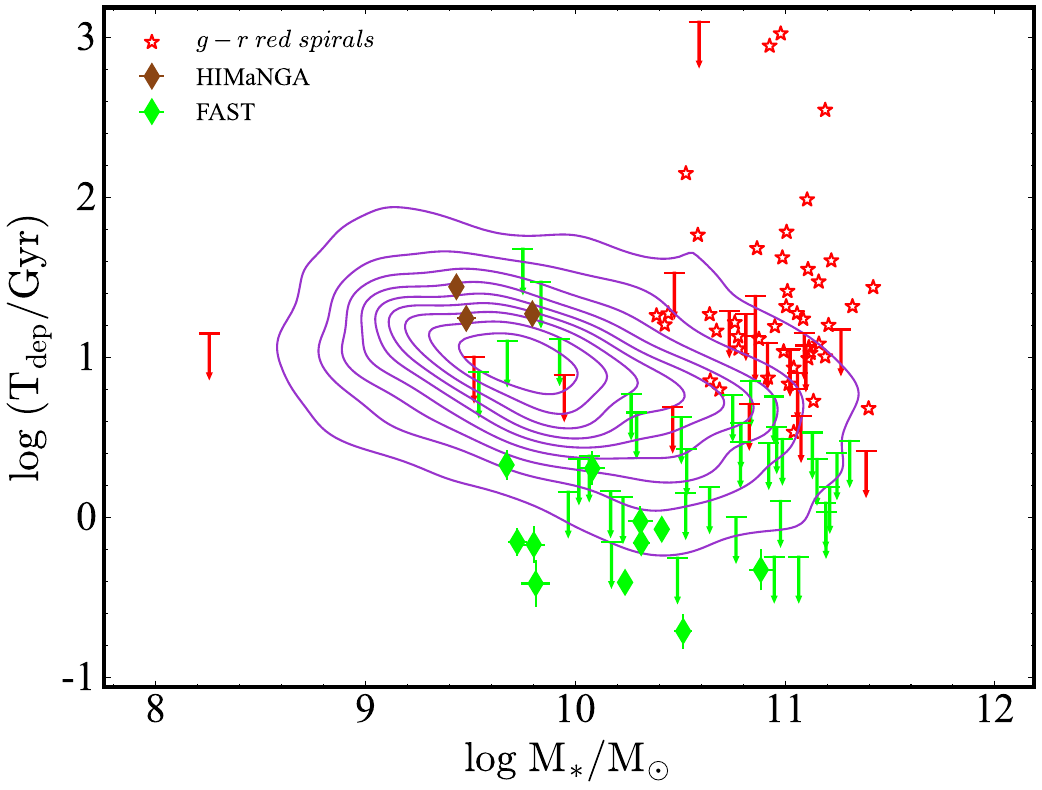}
      \label{fHI-Tdep}
     \end{minipage}
    }
\caption{H\,{\sc i} scale relations. a) $M_{\rm HI}$-$\Sigma_{1}$; The black dashed line represent the $\Sigma_{1}$ characteristic value \citep[][]{2024ApJ...968....3H}. b) $f_{\rm HI}$-$M_{\ast}$; The $f_{\rm HI}$ is defined as $M_{\rm HI}$/$M_{\ast}$. c) $f_{\rm HI}$-sSFR; The sSFR is defined as SFR/$M_{\ast}$. d) T$_{\rm dep}$-$M_{\ast}$; The T$_{\rm dep}$ is defined as $M_{\rm HI}$/SFR. The colors and labels are the same as Fig. \ref{MHI-mass}. The SFR and $M_{\ast}$ come from the fitting catalog of SED by \citet{2016ApJS..227....2S, 2018ApJ...859...11S}.}
\label{HI scale relation}
\end{figure}

In Fig. \ref{MHI-hist-mass}, we provided the histogram of $M_{\rm HI}$ and the $M_{\rm HI}$ - $M_{\ast}$ relation. The color scheme and labels are the same as Fig. \ref{structure properties}, where the numbers in different colors represent the median of samples. The filled diamonds represent detections from FAST (green) and H\,{\sc i}MaNGA (brown) for which the S/N for H\,{\sc i} $>$ 5, while the downward arrows denote the upper limits (S/N $\leq$ 5). The purple contour illustrates the number density distribution of 923 normal blue spirals with S/N for H\,{\sc i} $>$ 5, and the purple dashed line is their fitting result$^{\ref{foot9}}$. The KS test indicates that there is no significant statistical difference in the distribution of $M_{\rm HI}$ between star-forming S0s and normal blue spirals (p-value: 0.24), both showing significant differences from the distribution of $g - r$ red spirals, with p-values of 0.03 and 0.004, respectively. The larger $M_{\ast}$ ($>$ 10$^{10.4}$ $M_{\odot}$) of $g - r$ red spirals contributes to their larger $M_{\rm HI}$ (see Fig. \ref{MHI-mass}), consistent with \cite{2016MNRAS.455.2440M} and \cite{2021ApJ...918...53G}. In Figure \ref{MHI-mass}, we found that normal blue spirals indeed follow a tight H\,{\sc i} main sequence (slope: 0.40 $\pm$ 0.03, intercept: 5.73 $\pm$ 0.09, H\,{\sc i}MS), and the vast majority of star-forming S0s and $g - r$ red spirals are located below H\,{\sc i}MS \citep[e.g.,][]{2020MNRAS.493.1982J, 2021ApJ...918...53G}.

Next, we examined additional H\,{\sc i} scale relations. In Fig. \ref{HI scale relation}, the color scheme and labels are the same as in Fig. \ref{MHI-mass}. The H\,{\sc i} gas fraction ($f_{\rm HI}$) and the atomic gas depletion time (T$_{\rm dep}$) are defined as $f_{\rm HI}$ = $M_{\rm HI}$/$M_{\ast}$ and T$_{\rm dep}$ = $M_{\rm HI}$/SFR, respectively. We found no correlation between $M_{\rm HI}$ and $\Sigma_{1}$ as shown in Fig. \ref{MHI-sigma1}. Similar to Fig. \ref{sigma1-mass}, we also use the value of $\Sigma_{1}$ (10$^{9.4}$ $M_{\odot}$kpc$^{-2}$; black dashed line) provided by \citet{2024ApJ...968....3H} to identify galaxies with high-density cores. In Fig. \ref{MHI-sigma1}, $g - r$ red spirals and star-forming S0s exhibit $M_{\rm HI}$ comparable to normal blue spirals at a given $\Sigma_{1}$. However, when considering the $f_{\rm HI}$, we found that both groups are relatively gas-poor (see Fig. \ref{fHI-Mass}). The median values of $f_{\rm HI}$ are $-1.06$ ($g - r$ red spirals) and $-0.91$ (star-forming S0s), indicating that star-forming S0s have a slightly higher $f_{\rm HI}$. As the mass of the galaxy increases, the $f_{\rm HI}$ gradually decreases \citep[e.g.,][]{2016MNRAS.462.1749S}. The gas fractions of star-forming S0s and $g - r$ red spirals are relatively low, suggesting that galaxies with larger $M_{\ast}$ typically have lower $f_{\rm HI}$ \citep[e.g.,][]{2021ApJ...918...53G}.

\citet{2021ApJ...918...53G} found a correlation between $f_{\rm HI}$ and sSFR for SFGs and quiescent galaxies, confirming this viewpoint. For our targets, we also provided this relationship in Fig. \ref{fHI-sSFR}. Their relationship is essentially the Kennicutt-Schmidt law \citep[][]{1959ApJ...129..243S, 1998ApJ...498..541K}, although our measurements are globally average and the optical SF radii may be much smaller than the H\,{\sc i} disk sizes \citep[e.g.,][]{2021ApJ...918...53G}. In Fig. \ref{fHI-Tdep}, we illustrated the distributions of our samples on the T$_{\rm dep}$-$M_{\ast}$ plane. We found that the normal blue spirals have smaller atomic gas depletion time than $g - r$ red spirals. For the majority of $g - r$ red spirals, their T$_{\rm dep}$ exceed 10 Gyr, suggesting that H\,{\sc i} gas is not actively involved in the SF \citep[][]{2021ApJ...918...53G}. \citet{2017ApJS..233...22S} found an average T$_{\rm dep}$ around 0.65 $\pm$ 0.44 for galaxies on the star-forming main sequence using xGASS, which is consistent with our findings. However, the lower H\,{\sc i} content and higher average SFR in star-forming S0s result in relatively smaller T$_{\rm dep}$.

\section{Discussion}\label{discussion}

On average, $g - r$ red spirals are more compact than normal blue spirals and exhibit a steep SMR similar to quiescent S0s. This similarity between them suggests a potential evolutionary link, especially for massive galaxies \citep[e.g.,][]{2010MNRAS.405..783M, 2021ApJ...916...38Z, 2024MNRAS.528.2391C}. As the less pronounced spiral arms fade away and the residual SF in optically defined massive red spirals is exhausted, they may evolve into quiescent S0s \citep[e.g.,][]{2024MNRAS.528.2391C}. As SF decreases in normal blue spirals, their luminosity and size gradually decrease, leading them to evolve toward the red sequence \citep[][]{2011MNRAS.413..813C}. The SMR of star-forming S0s primarily overlaps with that of quiescent S0s and $g - r$ red spirals, deviating significantly from the SMR of normal blue spirals. The best-fit SMR for star-forming S0s is close to that of quiescent S0s and $g - r$ red spirals, suggesting similar gas dissipation scenarios in these populations \citep[e.g.,][]{2014ApJ...788...28V, 2024MNRAS.528.2391C}. Few normal blue spirals exhibit high $\Sigma_{1}$, indicating that a dense core is only a necessary condition for galaxy quenching, and they may be the rejuvenation of red spirals \citep[][]{2013ApJ...776...63F, 2020ApJ...897..162G, 2024ApJ...968....3H}. They may suffer from minor mergers or the accretion of fresh external gas \citep[e.g.,][]{2023A&A...678A..10G, 2024ApJ...968....3H}. For comparison, $g - r$ red spirals, quiescent S0s, and star-forming S0s have higher $\Sigma_{1}$ than normal blue spirals. Although there is a significant statistical difference in the distribution of $\Sigma_{1}$ among the four samples (p-value $<$ $1.46 \times 10^{-10}$), the proportion of dense cores in $g - r$ red spirals and quiescent S0s is significantly higher. Once a galaxy undergoes significant core growth (e.g., compaction process), it will lead to a rapid increase in $\Sigma_{1}$ and depletion of cold gas \citep[][]{2017ApJ...840...47B}, causing them to begin quenching and moving to the $\Sigma_{1}$-$M_{\ast}$ ridgeline \citep[][]{2013ApJ...776...63F} of quiescent population. This suggests an evolutionary sequence among galaxy types, where one galaxy may evolve into another over time.

The $g - r$ red spirals and quiescent S0s share similar old centers and higher [Mgb/Fe] (lower SF timescale), indicating the rapid formation of their bulges \citep[e.g.,][]{2024MNRAS.528.2391C} and similar formation mechanisms, while the residual SF in the outer disk of the former leads to their steeper profiles \citep[e.g.,][]{2021ApJ...916...38Z, 2024MNRAS.528.2391C}. In contrast, both radial profiles of star-forming S0s are relatively flat, positioned between those of normal blue spirals and quiescent S0s/$g - r$ red spirals. The relatively low [Mgb/Fe] of star-forming S0s compared to quiescent S0s and $g - r$ red spirals implies that they may have more extended SF processes. The $D_{\rm n}4000$ and [Mgb/Fe] profiles of normal blue spirals are also notably flat, consistent with their S$\acute{\rm e}$rsic index, where the lower values of [Mgb/Fe] reflect that the contribution of a more long-term SF history \citep[][]{2024MNRAS.528.2391C}. The different SF timescale between $g - r$ red spirals and normal blue spirals indicates that they may have different formation mechanisms \citep[][]{2019ApJ...883L..36H, 2024MNRAS.528.2391C}. There is no significant statistical difference in halo mass distributions between $g - r$ red spirals and quiescent S0s (p-value: 0.09), and both groups predominantly exceed the critical mass. The halo mass distribution of star-forming S0s shows no significant difference statistically from that of quiescent S0s and $g - r$ red spirals, with over 82$\pm$5$\%$ of these galaxies above the critical mass threshold. It is generally believed that if the halo mass of a galaxy exceeds this critical value, the galaxy will be quenched \citep[e.g.,][]{2014MNRAS.440..889S, 2021ApJ...918...53G, 2024ApJ...963...86L}. In terms of H\,{\sc i} content, $g - r$ red spirals generally have larger $M_{\rm HI}$ due to their larger $M_{\ast}$ \citep[][]{2016MNRAS.455.2440M, 2021ApJ...916...38Z}. Normal blue spirals indeed follow the tight H\,{\sc i}MS \citep[][]{2020ApJ...897..162G, 2020MNRAS.493.1982J}, whereas most star-forming S0s and $g - r$ red spirals lie below the H\,{\sc i}MS. The residual gas in $g - r$ red spirals sustains limited ongoing SF, likely confined to the disk \citep[][]{2020ApJ...897..162G, 2021ApJ...916...38Z, 2024MNRAS.528.2391C}. Moreover, there is no significant correlation between the $M_{\rm HI}$ and $\Sigma_{1}$, indicating that the dense core is only a necessary condition for galaxy quenching \citep[e.g.,][]{2020ApJ...897..162G, 2024ApJ...968....3H}. Compared to normal blue spirals, both star-forming S0s and $g - r$ red spirals are relatively gas-poor. As $M_{\ast}$ of the galaxy increases, $f_{\rm HI}$ declines \citep[e.g.,][]{2016MNRAS.462.1749S}. The relationship between $f_{\rm HI}$ and sSFR may be linked to the Kennicutt-Schmidt law \citep[][]{2008AJ....136.2846B}. Variations in SFRs and $f_{\rm HI}$ across galaxies influence their T$_{\rm dep}$.

The $g - r$ red spirals share many similarities with quiescent S0s, including structures (SMR, $\Sigma_{1}$, and S$\acute{\rm e}$rsic index), SPs properties, and galaxy halo mass. Simulations suggest that the spiral arms are transient and recurrent \citep[e.g.,][]{2013ApJ...763...46B, 2013ApJ...766...34D}, potentially reappearing during the formation of S0s \citep[e.g.,][]{2020MNRAS.498.2372D}. \citet{2010MNRAS.405..783M} provided three possible origins of red spirals: old spirals that have used up fuel, the transformation of normal blue spirals that have undergone some processes, or the evolution of normal blue spirals due to bar instability. Furthermore, previous studies on red spirals have also shown that they may be one of the channels for the formation of S0s \citep[][]{2018MNRAS.474.1909F, 2019ApJ...880..149P}, which applies to S0s with various origin mechanisms \citep[][]{2024MNRAS.528.2391C}. Despite differences in the definition of red spirals across studies, the primary distinction lies in the presence of residual SF in their outer regions \citep[][]{2021ApJ...916...38Z}. 

In light of these observations, we discussed possible evolutionary scenarios (Sections \ref{sec4.1} and \ref{sec4.2}) of star-forming S0s, illustrated in Fig. \ref{evlolution path}. Numerical markers in the figure indicate the stages a galaxy might traverse during its evolution between states.

\subsection{Experiencing Red Spirals Transient}\label{sec4.1}

In isolated or low-density environments, red spirals may simply be old spiral galaxies that have exhausted their fuel \citep[e.g.,][]{2010MNRAS.405..783M, 2019ApJ...880..149P}. This gas consumption may be accelerated through internal processes related to the bar structure \citep[e.g.,][]{2010MNRAS.405..783M, 2018MNRAS.474.1909F, 2019ApJ...880..149P}. In contrast, in high-density environments, surrounding interactions or gas-stripping processes can transform normal blue spirals into red spirals without disrupting their spiral structures, thus halting SF \citep[e.g.,][]{2010MNRAS.405..783M, 2019ApJ...880..149P}. During the transition from normal blue spirals, certain disruptive processes may prevent low-mass galaxies from being observed as red spirals, contributing to the bending of red spirals in the SMR \citep[][]{2010MNRAS.405..783M}. When the less pronounced spiral arms and residual SF in red spirals disappeared, they evolved into the quiescent S0s \citep[e.g.,][]{2002ApJ...577..651B, 2024MNRAS.528.2391C}. Quiescent S0s tend to have a dense environment \citep[e.g.,][]{2022A&A...659A..46B} and may experience external gas accretion or minor mergers that reignite SF, leading them to evolve into star-forming S0s \citep[e.g.,][]{2015A&A...583A.103G, 2023A&A...678A..10G}. This scenario aligns with previously proposed explanations for the bending observed in ETGs on the SMR \citep[][]{2013MNRAS.432.1862C}. More recently, \citet{somawanshi2024understanding} found [$\alpha$/Fe] dichotomy in edge-on S0 galaxy - ESO 544-27 - with its thin and thick discs dominated by low and high [$\alpha$/Fe] SPs respectively, and also discovered the metal-rich younger SPs ($<$ 2 Gyr), indicating that it was nearly quenched until its SF was reignited recently first in the outer and inner thick disc and then in the thin disc. This scenario can explain the lower levels of $D_{\rm n}4000$ and [Mgb/Fe] and is similar to the rejuvenation of SF in red spirals, resulting in the observation of normal blue spirals with high $\Sigma_{1}$ \citep[e.g.,][]{2020ApJ...897..162G, 2024ApJ...968....3H}. As SF in star-forming S0s gradually quenches, they are expected to transition back to quiescent S0s (Fig. \ref{path1}).

Star-forming S0s and red spirals also exhibit similarities in various aspects, including structures (SMR, $\Sigma_{1}$), galaxy halo mass, and $f_{\rm HI}$. We suspect that star-forming S0s may be a special phase between red spirals and quiescent S0s (Fig. \ref{path2}). When a red spiral galaxy experiences external fresh gas supply or merging \citep[][]{2013MNRAS.432.1862C, 2020ApJ...897..162G, 2024ApJ...968....3H}, there will be the rejuvenation of SF. As the less pronounced spiral arms fade away and the residual SF is exhausted, the star-forming S0s will evolve into quiescent S0s. Similar to Fig. \ref{path1}, quiescent S0s can also return to star-forming S0s \citep[e.g.,][]{2015A&A...583A.103G, 2023A&A...678A..10G}. There is a slight increase of gas ($f_{\rm HI}$) in star-forming S0s. We found that red spirals are mostly concentrated at the high-mass end (log $M_{\ast}$ $>$ 10.5 $M_{\odot}$), while the mass of other galaxies covers a wider range. We suggest that these two situations (Figures \ref{path1} and \ref{path2}) may only apply to massive galaxies.

\subsection{Evolved directly from Normal Blue Spirals}\label{sec4.2}

\begin{figure}[!ht]
 \centering
    \subfigure[]
    {
     \begin{minipage}{8.2cm}
      \centering
      \includegraphics[scale=0.45]{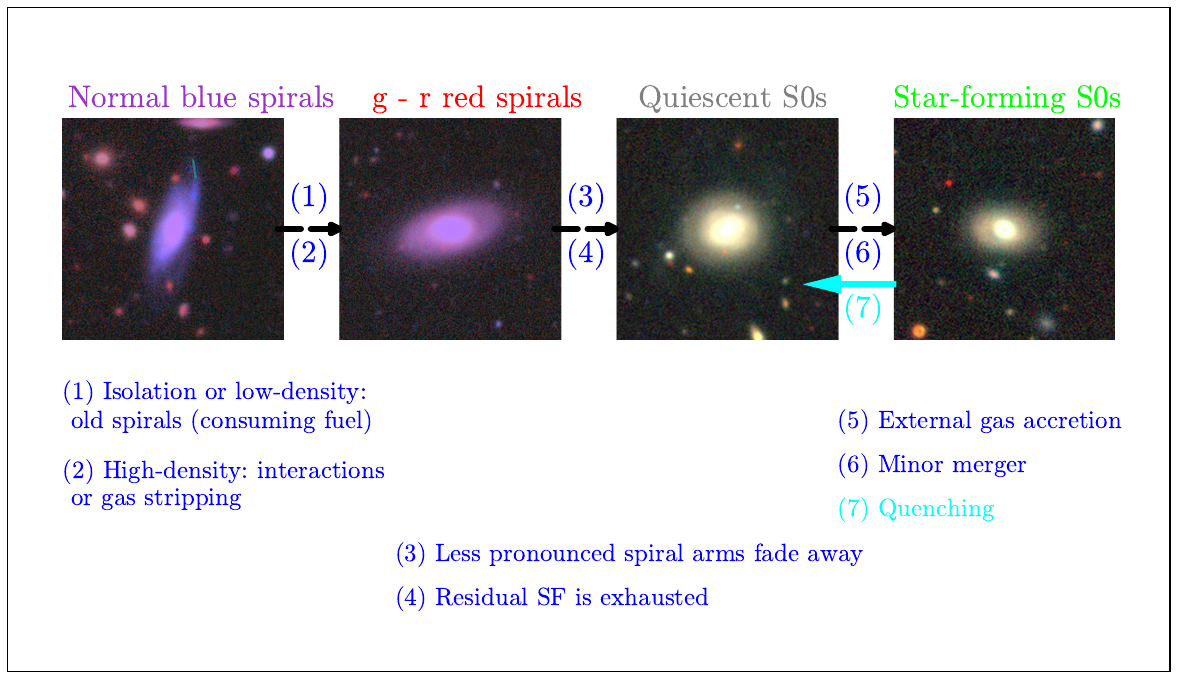}
      \label{path1}
     \end{minipage}
    }
    \subfigure[]
    {
     \begin{minipage}{8.2cm}
      \centering
      \includegraphics[scale=0.45]{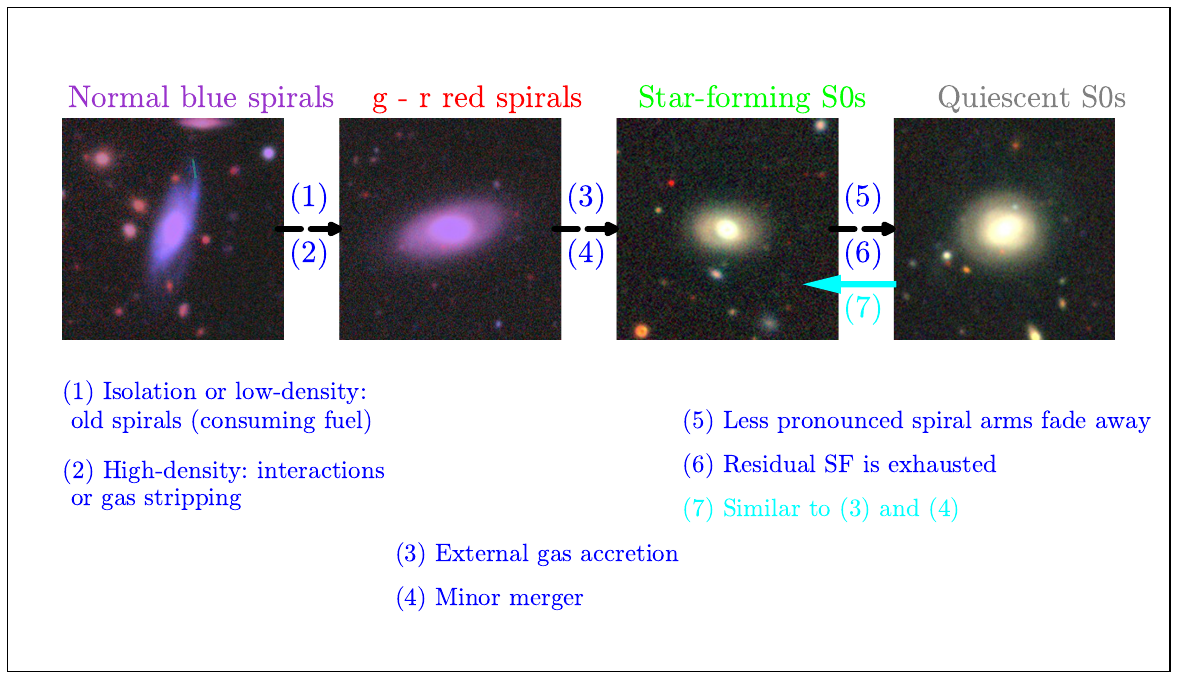}
      \label{path2}
     \end{minipage}
    }
    \subfigure[]
    {
     \begin{minipage}{8.2cm}
      \centering
      \includegraphics[scale=0.45]{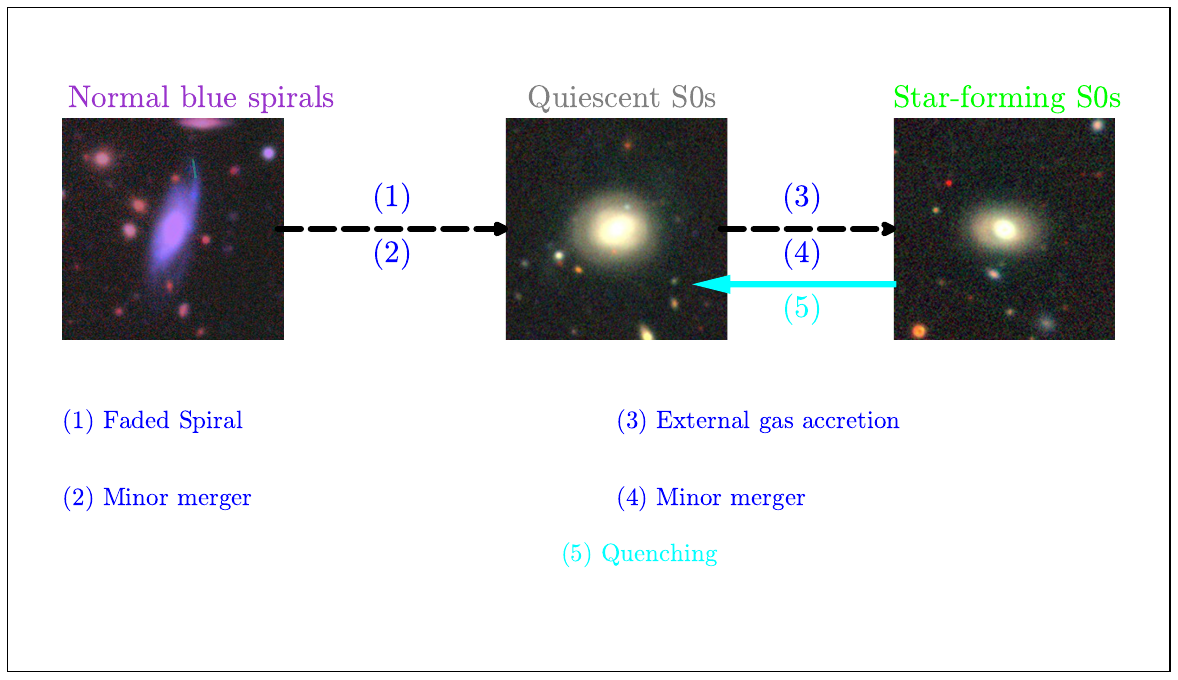}
      \label{path3}
     \end{minipage}
    }
    \subfigure[]
    {
     \begin{minipage}{8.2cm}
      \centering
      \includegraphics[scale=0.45]{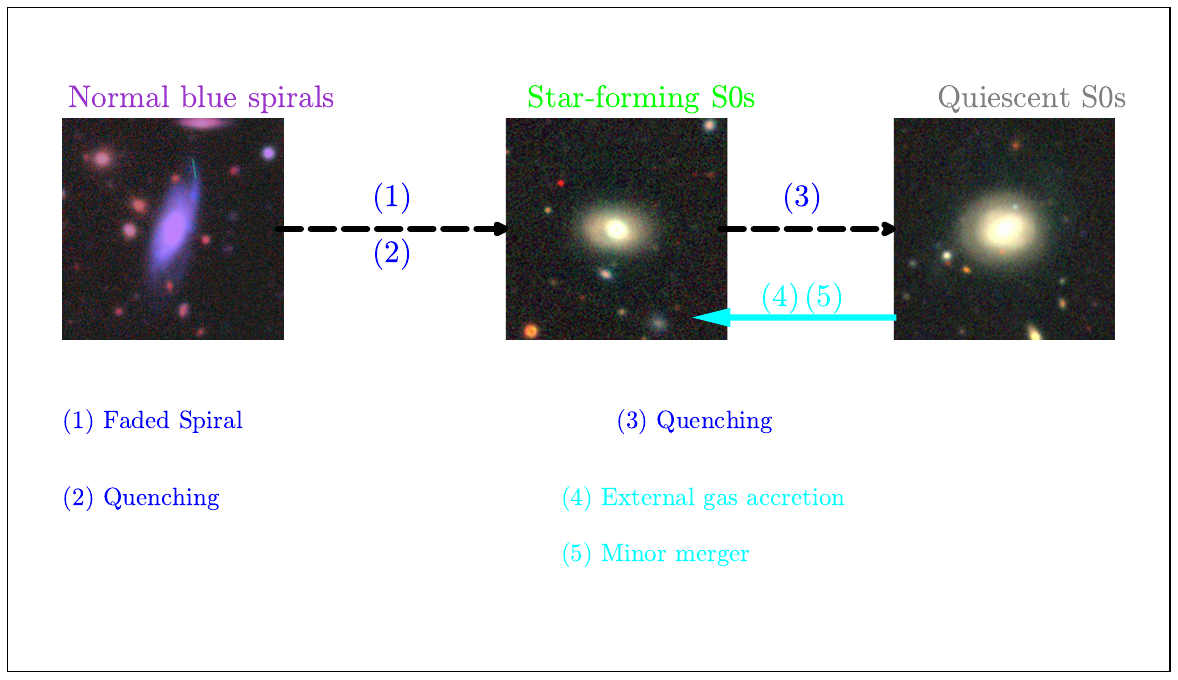}
      \label{path4}
     \end{minipage}
    }
\caption{A schematic representation of the possible evolutionary pathways of our sample. a) normal blue spirals - $g - r$ red spirals - quiescent S0s - star-forming S0s; b) normal blue spirals - $g - r$ red spirals - star-forming S0s - quiescent S0s; c) normal blue spirals - quiescent S0s - star-forming S0s; d) normal blue spirals - star-forming S0s - quiescent S0s. In this image, we use g-/r-/z-bands false-color images from DESI \citep[][]{2019AJ....157..168D} DR10 to represent the corresponding galaxy types. The process that a galaxy may undergo during its evolution between two states is indicated by numerical markers. The color coding is the same as in Fig. \ref{structure properties}.}
\label{evlolution path}
\end{figure}

Red spirals are rare and unique \citep[e.g.,][]{2010MNRAS.405..783M}, and it remains uncertain whether galaxies consistently pass through such specialized stages during evolution. The special population, star-forming S0s, has diverse and complex formation mechanisms \citep[e.g.,][]{2020MNRAS.498.2372D, 2021MNRAS.508..895D, 2022MNRAS.509.1237X, 2024A&A...691A.107C}, and may not experience the intermediate state of red spirals during their formation process (Figures \ref{path3} and \ref{path4}). Specifically, star-forming S0s are either the rejuvenation of SF in quiescent S0s \citep[Fig. \ref{path3};][]{2015A&A...583A.103G, 2023A&A...678A..10G}, or a special phase between normal blue spirals and quiescent S0s \citep[Fig. \ref{path4}; e.g.,][]{2022A&A...659A..46B}. Both pathways bypass the red spiral stage and are not constrained by galaxy mass. However, the ``fate" of blue ETGs - star-forming S0s is still uncertain. This proposed scheme is based on limited observational data; additional insights from molecular gas data, deeper surveys, and more advanced simulations will be essential for refining our understanding of these evolutionary paths.

\section{Conclusions}\label{conclusions}

In this work, we compared star-forming S0s with other galaxies in terms of structures (SMR, $\Sigma_{1}$, and S$\acute{\rm e}$rsic index), SPs properties ($D_{\rm n}4000$ and [Mgb/Fe]), galaxy halo mass, and H\,{\sc i} content. Based on these observational results, we discussed the possible positions of star-forming S0s in the evolutionary sequence. Our main conclusions are as follows:

1). We found that the SMRs of $g - r$ red spirals and star-forming S0s are between normal blue spirals and quiescent S0s, but they have a steeper slope similar to quiescent S0s, and both exhibit bending phenomena (see Sect. \ref{global-properties}). This result suggests that they may have similar gas dissipation scenarios. Moreover, they all have higher $\Sigma_{1}$ compared to normal blue spirals. The emergence of normal blue spirals with high $\Sigma_{1}$ also indicates that a dense core is only a necessary condition for galaxy quenching, and they may be the rejuvenation of red spirals \citep[][]{2024ApJ...968....3H}. The proportion of dense cores (log $\Sigma_{1}$ $>$ 9.4) in $g - r$ red spirals and quiescent S0s is significantly higher. There is no significant statistical difference in the distribution of S$\acute{\rm e}$rsic index between $g - r$ red spirals and quiescent S0s (p-value: 0.37), which are different from normal blue spirals and star-forming S0s. The vast majority of normal blue spirals (79$\pm$5$\%$) have S$\acute{\rm e}$rsic index $<$ 2.0.

2). We examined the radial profiles of the spectral indices ($D_{\rm n}4000$ and [Mgb/Fe]). Both $g - r$ red spirals and quiescent S0s exhibit similar old centers and higher [Mgb/Fe], but the former has steeper profiles. The presence of residual SF in the outer regions of red spirals leads to such results. The high [Mgb/Fe] of centers between the two indicates the rapid formation of bulges. The radial profile gradients of $D_{\rm n}4000$ and [Mgb/Fe] for star-forming S0s are relatively flat, but they are both between normal blur spirals and quiescent S0s/$g - r$ red spirals. Furthermore, considering the S$\acute{\rm e}$rsic index of the vast majority of normal blue spirals and their more extended SF, we can understand their very flat $D_{\rm n}4000$ and [Mgb/Fe] profiles, and the lower [Mgb/Fe] indicates the contribution of a more long-term SFH.

3). We found that over 82$\pm$5 $\%$ of galaxies in star-forming S0s have halo mass exceeding the critical mass (log$M_{\rm halo}$ = $10^{12}$ $M_{\odot}/h$) for galaxy quenching, and there is no significant statistical difference between their halo mass distribution and that of quiescent S0s (p-value: 0.99) and $g - r$ red spirals (p-value: 0.46). The distributions of halo mass for $g - r$ red spirals and quiescent S0s also show no significant statistical difference (p-value: 0.09), and the vast majority of them ($\geq$ 83.00$\%$) exceed the critical mass. Other statistical tests (AD test and Permutation test) have also shown the same results (Table \ref{test_result}). In contrast, the smaller galaxy halo mass in normal blue spirals has a higher fraction (36$\pm$6$\%$).

4). We used FAST to observe H\,{\sc i} in 41 star-forming S0s for 10 hours, detecting significant H\,{\sc i} emission in 11 of them, giving a detection rate of approximately 27$\%$. As a supplement, we collected 10 additional sources with H\,{\sc i} detection (S/N of H\,{\sc i} $>$ 3) provided by H\,{\sc i}MaNGA. $g - r$ red spirals seem to have larger $M_{\rm HI}$, but this should be attributed to their larger mass ($>$ $10^{10.4}$ $M_{\odot}$). Consistent with previous studies, normal blue spirals indeed follow a tight H\,{\sc i}MS, and the $f_{\rm HI}$ of all samples decreases with increasing $M_{\ast}$. The differences in star-forming states and $f_{\rm HI}$ among different samples lead to variations in T$_{\rm dep}$, which align with the previous. Notably, both $g - r$ red spirals and star-forming S0s are relatively gas-poor. Moreover, our samples do not show any significant correlation on the $M_{\rm HI}$ - $\Sigma_{1}$ plane.

5). Normal blue spirals in different environments undergo different processes to transform into $g - r$ red spirals, and when the less pronounced spiral arms and residual SF in the latter disappear, they evolve into quiescent S0s. Subsequently, possible external gas accretion and minor merger reignited the SF in quiescent S0s, causing them to become star-forming S0s. Of course, considering the similarities between star-forming S0s and $g - r$ red spirals in many aspects, we suspect that star-forming S0s may also be a special phase between $g - r$ red spirals and quiescent S0s. Most $g - r$ red spirals are located at the high-mass end, so these two situations may only apply to massive star-forming S0s. In addition, it is also possible that star-forming S0s may not experience the red spiral state during their evolution process. They can be the rejuvenation of SF in quiescent S0s or a special phase between normal blue spirals and quiescent S0s. The above sequences we have discussed are only based on all observation results. We look forward to more data (e.g., molecular gas data), deeper surveys, and more advanced simulations in the future.

\vspace{5mm}
We are grateful to the anonymous referee for her/his thoughtful review and very constructive suggestions, which greatly improved this paper. We thank the Statistics Editor for suggesting more advanced tests. We thank Dr. Qiusheng Gu, Dr. Cheng Li, Dr. Cheng Cheng, Dr. Renbin Yan, Dr. Cai-Na Hao, Dr. Rui Guo, Dr. Jie Wang, Dr. Lan Wang, Dr. Yingjie Jing, and Dr. Yinghe Zhao for the helpful discussion. This work made use of the data from FAST (Five-hundred-meter Aperture Spherical Radio Telescope).  FAST is a Chinese national mega-science facility, operated by the National Astronomical Observatories, Chinese Academy of Sciences. J.W. acknowledges the National Key R$\&$D Program of China (grant No.2023YFA1607904) and the National Natural Science Foundation of China (NSFC) grants 12033004, 12221003, 12333002 and the science research grant from CMS-CSST-2021-A06. This work is supported by the CNSA program D050102. T.C. acknowledges the China
Postdoctoral Science Foundation (grant No. 2023M742929),
and the NSFC grants 12173045 and 12073051. X.X. acknowledges the NSFC grant 12403018, the China Postdoctoral Science Foundation (grant No. 2023M741639), and the Jiangsu Funding Program for Excellent Postdoctoral Talent. The research made use of Photutils, an Astropy package for detection and photometry of astronomical sources \citep[][]{2022zndo...7419741B}.

Funding for the Sloan Digital Sky Survey IV has been provided by the Alfred P. Sloan Foundation, the U.S. Department of Energy Office of Science, and the Participating Institutions. SDSS-IV acknowledges support and resources from the Center for High-Performance Computing at the University of Utah. The SDSS website is \url{www.sdss.org}. SDSS-IV is managed by the Astrophysical Research Consortium for the Participating Institutions of the SDSS Collaboration including the Brazilian Participation Group, the Carnegie Institution for Science, Carnegie Mellon University, the Chilean Participation Group, the French Participation Group, Harvard-Smithsonian Center for Astrophysics, Instituto de Astrof\'{i}sica de Canarias, The Johns Hopkins University, Kavli Institute for the Physics and Mathematics of the Universe (IPMU) / University of Tokyo, Lawrence Berkeley National Laboratory, Leibniz Institut f\"{u}r Astrophysik Potsdam (AIP), Max-Planck-Institut f\"{u}r Astronomie (MPIA Heidelberg), Max-Planck-Institut f\"{u}r Astrophysik (MPA Garching), Max-Planck-Institut f\"{u}r Extraterrestrische Physik (MPE), National Astronomical Observatory of China, New Mexico State University, New York University, University of Notre Dame, Observat\'{o}rio Nacional / MCTI, The Ohio State University, Pennsylvania State University, Shanghai Astronomical Observatory, United Kingdom Participation Group, Universidad Nacional Aut\'{o}noma de M\'{e}xico, University of Arizona, University of Colorado Boulder, University of Oxford, University of Portsmouth, University of Utah, University of Virginia, University of Washington, University of Wisconsin, Vanderbilt University, and Yale University.

\vspace{5mm}
\facilities{Sloan; FAST}
\software{astropy \cite[][]{2013A&A...558A..33A, 2018AJ....156..123A, 2022ApJ...935..167A}; Tool for OPerations on Catalogues And Tables \citep[TOPCAT;][]{2005ASPC..347...29T}}

\bibliography{sample631}{}
\bibliographystyle{aasjournal}


\end{CJK*}
\end{document}